\documentclass[10pt,a4paper]{article}

\usepackage{hyperref} % for better urls
\usepackage[T1]{fontenc}
\usepackage[utf8]{inputenc}
\usepackage{authblk}
\usepackage{graphicx}
\usepackage{amsmath}
\usepackage[left=2cm,right=2cm,top=2cm,bottom=2cm]{geometry}

\title{Modeling cell crawling strategies with a bistable model: From amoeboid to fan-shaped cell motion}
\author[1]{Eduardo Moreno\thanks{eduardo.moreno.ramos@upc.edu}}
\author[2]{Sven Flemming\thanks{svenflem@uni-potsdam.de}}
\author[1,4]{Francesc Font\thanks{ffont@crm.cat}}
\author[3]{Matthias Holschneider\thanks{hols@uni-potsdam.de}}
\author[2,5]{Carsten Beta\thanks{beta@uni-potsdam.de}}
\author[1]{Sergio Alonso\thanks{s.alonso@upc.edu}}
\affil[1]{Department of Physics, Universitat Polit\`ecnica de Catalunya, 08028 Barcelona, Spain.}
\affil[2]{Institute of Physics and Astronomy, University of Potsdam, 14476 Potsdam, Germany.}
\affil[3]{Institute of Mathematics, University of Potsdam, 14476 Potsdam, Germany.}
\affil[4]{Centre de Recerca Matem\`{a}tica, Campus de Bellaterra, Edifici C, Bellaterra 08193, Barcelona, Spain.}
\affil[5]{Max Planck Institute for Dynamics and Self-Organization, 37077 G\"ottingen, Germany.}

\begin{document}
		\maketitle
		\begin{abstract}
			Eukaryotic cell motility involves a complex network of interactions between biochemical components and mechanical processes.
			The cell employs this network to polarize and induce shape changes that give rise to membrane protrusions and retractions, ultimately leading to locomotion of the entire cell body.
			The combination of a nonlinear reaction-diffusion model of cell polarization, noisy bistable kinetics, and a dynamic phase field for the cell shape permits us to capture the key features of this complex system to investigate several motility scenarios, including amoeboid and fan-shaped forms as well as intermediate states with distinct displacement mechanisms.
			We compare the numerical simulations of our model to live cell imaging experiments of motile {\it Dictyostelium discoideum} cells under different developmental conditions.
			The dominant parameters of the mathematical model that determine the different motility regimes are identified and discussed.
			%Satisfactory qualitative and in some cases quantitative agreement between our simulations and experiments is found.
		\end{abstract}
		
		
		\section{Introduction}
		
		The biochemical and biophysical mechanisms involved in cell motility have been extensively studied during the past years.
		They are among the most intriguing problems in cell biology, ranging from single cells to multicellular organisms. 
		Before the cell begins to move, it has to define a front and a back to specify an axis of propagation.
		This process is known as cell polarization~\cite{mogilner2012cell}. 
		It sets the direction in which protrusions are formed that drive the cell forward.
		
Cell locomotion has been extensively studied using keratocytes, which move in a highly persistent fashion and adopt a characteristic fan-like shape \cite{mogilner_experiment_2019}.
Also neutrophils have been intensely investigated.
They display a less persistent movement with more frequent random changes in direction that is known as amoeboid motility~\cite{haastert_chemotaxis_2004}.
A well-established model system to study actin-driven motility in eukaryotic cells is the social amoeba {\it Dictyostelium discoideum} ({\it D. discoideium})~\cite{annesley_dictyostelium_2009}.
The cells of this highly motile single-celled microorganism typically display pseudopod-based amoeboid motility but also other forms, such as blebbing motility or keratocyte-like behavior have been observed.

Many aspects of cell motility such as cytoskeletal mechanics~\cite{goehring2013cell}, intracellular signaling dynamics~\cite{beta2017intracellular,rappel2017mechanisms}, or membrane deformation~\cite{allard2013traveling} have been modeled using mathematical and computational methods. 
Cell polarity formation, which is a key features of motility mechanisms to determine the front and back of the cell, often shows bistable dynamics.
A reaction-diffusion system with bistable kinetics is thus a common choice to model the intracellular polarity dynamics~\cite{jilkine2011comparison}.
%Although in certain cells polarity is a dynamical process which can, for example, periodically change with time due to internal actin waves \cite{beta2017intracellular}, the mechanism of cell polarity seems to be related to a bistable condition, determining the front and the back of the cell. 
Bistable conditions of an intracellular dynamical process can be obtained by a mass-controlling mechanism between the cytosolic and membrane attached concentrations of biochemical components~\cite{mori2008wave,otsuji2007mass}.
This may be relevant at different levels of the cytoskeleton, for example, when different forms of actin are involved~\cite{beta2008bistable,schroth-diez_propagating_2009,beta2010bistability} or at the level of the related signaling pathways, involving phospholipids and enzymes at the cell membrane~\cite{matsuoka2018mutual,altschuler2008spontaneous}.
Cell polarity may be also induced by an external chemical gradient~\cite{iglesias2008navigating}.

There are several mathematical tools to simultaneously model the pattern formation process inside the cell and the dynamics of the cell border, which is required to obtain a full description of a crawling cell.
One of the most commonly employed methods to model such a free-boundary problem is to introduce an additional phase field, which is one inside and zero outside the cell and keeps the correct boundary conditions while the borders are moving~\cite{kockelkoren2003computational}, even in the limit of the sharp interface between the interior and the exterior of the cell~\cite{camley2013periodic}.
The first attempts to employ a phase field modeling to study cell locomotion where applied to keratocyte motility~\cite{shao2010computational,ziebert2011model,shao2012coupling} because the persistence of motion of these cells facilitates the implementation of the model.
These models have also been extended to discuss, for example, the rotary motion of keratocytes~\cite{Camley17} and the interactions among adjacent cells~\cite{lober2014modeling}. 

Later, the use of the phase field has been extended to model other generic properties of the moving cells \cite{najem2013phase,kulawiak2016modeling} and, in particular, has been also employed to describe the random motion of amoeboid cells, such as neutrophils or {\it D. discoideum}.
The phase field approach has been employed to model the viscoelasticity of the cell~\cite{moure2016computational,moure2017phase}, the effect of biochemical waves in the interior of the cell~\cite{taniguchi2013phase}, as well as wave-induced cytofission of cells~\cite{flemming2019}. 
%Sven: I rewrote the following section to be more precise regarding the cell lines used. The original section is below this new part (I used many parts of your text)
The random motion of the cell requires a stochastic bistable process in combination with a phase field~\cite{Alonso18} to successfully recover the fluctuating displacements and shape deformations.
Such a stochastic bistable model is able to capture the cell-to-cell variability observed in the motion patterns of amoeboid {\it D. discoideum} cells by tuning a single model parameter~\cite{Alonso18}.
However, {\it D. discoideum} cells are known to show a more diverse spectrum of motility modes, for example when certain genes are knocked out, when phosphoinositide levels are artificially altered, or under specific developmental conditions.
This includes a phenotype which is reminiscent of keratocyte motility, where the cell adopts a fan-like shape and moves persistently, perpendicular to the elongated axis of the cell body -- the so-called fan-shaped phenotype -- and a form where cells adopt a pancake-like shape, moving erratically without a clear direction of polarization~\cite{asano2004keratocyte,miao2017altering,cao2019plasticity}. 

Here, we perform a systematic analysis of a previously introduced model that is based on a stochastic bistable reaction-diffusion system in combination with a dynamic phase field~\cite{Alonso18}.
Such phenomenological model may provide a better understanding of how to relate the experimental parameters to specific cellular behaviors, because cell-to-cell variability often masks such relation.
Along with the model, we analyze experimental data of a non-axenic {\it D. discoideum} wildtype cell line (DdB) that carries a knockout of the RasGAP homologue NF1 (DdB NF1 null cells).
In this cell line, amoeboid and fan-shaped cells are observed, depending on the developmental conditions.
%along with several transition states between these phenotypes.
A detailed comparison of the experimental data to simulations of the stochastic bistable phase field model is presented.
By tuning the intensity of the noise and the area covered by the bistable field, the model simulations recover similar motility phenotypes as observed in experiments, ranging from highly persistent fan-shaped cells to standard amoeboid motion.
Furthermore, the simulations predict intermediate unstable states and also a transition from straight to rotary motion of the fan-shaped cells.
These forms of motility that have so far been neglected in {\it D. discoideum}, were also observed in the experimental data and are systematically studied in the framework of our mathematical model.

\section{Materials and Methods}

\subsection{Experimental Methods}

All experiments were performed with non-axenic \textit{D.~discoideum} DdB NF1 KO cells \cite{bloomfield2015neurofibromin}, which were cultivated in 10~cm dishes with Sorensen's buffer (8~g KH$_2$PO$_4$, 1.16~g Na$_2$HPO$_4$, pH 6.0) supplemented with 50~$\mu$M MgCl$_2$, 50~$\mu$M CaCl$_2$ and \textit{Klebsiella aerogenes} at an OD$_{600}$ of 2. The cells expressed Lifeact-GFP via the episomal plasmid SF99, which is based on a new set of vectors for gene expression in non-axenic \textit{D.~discoideum} strains \cite{paschke2018rapid}. Plasmids were transformed as described before \cite{paschke2018rapid} with an ECM2001 electroporator using three square wave pulses of 500~V for 30~ms in electroporation cuvettes with a gap of 1 mm. G418 (5 $\mu$g/ml) and Hygromycin (33 $\mu$g/ml) were used as selection markers.

The phenotype of DdB NF1 KO cells differs between individual cells of a population and especially between different developmental stages. When cultivated in buffer supplemented with bacteria, cells are in the vegetative state and the predominant phenotype is amoeboid with very little movement due to the abundance of bacteria. After several hours of starvation, cell enter the developed state and the probability to observe a fan-shaped phenotype is increased. Preparation of the cells forexperiments therefore differed between experiments. Cells were washed to removethe bacteria and (i) suspended in Sörensen’s Buffer immediately after washing to obtain mainly amoeboid cells with high motility or (ii) starved for 3-6 hours to obtain a high percentage of fan-shaped cells. After starvation, cells were seeded in microscopy dishes at low density for imaging. Usually in the beginning of an experiment, many cells showed the amoeboid or the intermediate phenotype with regular switches from amoeboid to fan-shaped motility and vice versa. The percentage of fan-shaped cells increased over time and the fan-shaped phenotype became more stable. An increase in the number of fan-shaped cells during
development has also been described for the D. discoideum Ax2 AmiB knockout strain (Asano et al., 2004). Note however, that the effects of cell development on the phenotype of DdB NF1 KO cells showed a high day-to-day variability and best results were accomplished with fresh K. aerogenes cells. The cells were transferred to a 35 mm glass bottom microscopy dish (FluoroDish, World Precision Instrumnets) and diluted to a concentration enabling imaging of single cells. For imaging an LSM 780 (Zeiss, Jena) with a 488 nm argon laser and a 63x or a 40x oil objective lens were used.

\subsection{Computational Model}

We investigate different types of cell motility based on a minimal model that couples a concentration field accounting for the complex biochemical reactions occurring in the interior of the cell to an auxiliary phase field describing the evolution of the cell shape.
The use of a phase field is a well-established approach to deal with problems of evolving domains/geometries without the need of explicitly tracking the domain boundaries, which has been exploited to tackle moving boundary problems of different nature such as crack propagation \cite{boettinger2002phase}, solidification \cite{pons2010helical}, or fluid interface motion \cite{folch1999phase}.
In particular, the versatility of this approach has been used for modelling cell-shape evolution and locomotion \cite{shao2012coupling,taniguchi2013phase,Camley17}. 
The model we use here has been previously introduced in~\cite{Alonso18} and will be summarized below.

In what follows we will consider the dynamics of a generic activatory biochemical component at to the substrate-attached cell-membrane and, thus, restrict ourselves to an idealized 2D geometry. The phase field $\phi(\boldsymbol{x},t)$ smoothly varies between the values of $\phi=1$ inside and $\phi=0$ outside the cell, respectively. The phase field allows to implicitly impose no-flux boundary conditions at the cell border, which we assume to be where the phase field takes the value of $\phi=0.5$. Following the work by Shao \textit{et\,al.}~\cite{shao2012coupling}, the phase field evolves according to the equation 
\begin{equation}\label{pf}
\tau \frac{\partial \phi}{\partial t} = \gamma \left(\nabla^2 \phi -\frac{G'(\phi)}{\epsilon^2}\right)
- \beta \left(\int \phi\, dA - A_0 \right)\left| \nabla \phi \right| + 
\alpha\, \phi\, c \left| \nabla \phi \right|  \,, 
\end{equation}
where $G(\phi) = 18\,\phi^2\,(1-\phi)^2$ is a double well potential. The phase field equation is the result of a force balance involving forces of different nature acting on the cell body. The terms with $| \nabla \phi|$ affect the border of the cell, while the others affect the volume. The first term on the right hand side of eq.~\eqref{pf} corresponds to the surface energy of the cell membrane, where $\gamma$ is the surface tension (note that value is obtained assuming a cell height of 0.15 $\mu$m \cite{shao2012coupling}) and $\epsilon$ the mathematical definition of the width of the cell boundary. The second term ensures that the cell area is kept close to $A_0$. The last term represents the active force of the biochemical field $c(\boldsymbol{x},t)$ on the cell membrane. The parameters that control the impact of the area conservation constraint and the active force, $\beta$ and $\alpha$, respectively, are kept constant in our simulations. The term on the left hand side of eq.~\eqref{pf} accounts for cell-substrate friction. 

A complete derivation of eq.~\eqref{pf} can be found in \cite{shao2010computational}.

Cell polarization is implemented in our model by assuming noisy bistable dynamics resulting from a non-linear reaction-diffusion equation for the concentration $c(\boldsymbol{x},t)$. The biochemical field $c(\boldsymbol{x},t)$ represents a dimensionless generic concentration variable that accounts for different subcellular components that promote the growth of filamentous actin (F-actin) such as active Ras, PI3K or PIP$_3$ \cite{swaney2010eukaryotic}. It is related to the intensity of the Lifeact-GFP marker for F-actin in our experiments. 

Imaging experiments with {\it D. discoideum} typically show rich dynamical patterns in the cell cortex and at the cell membrane. Deriving a detailed model that captures the full complexity of the underlying biochemical reactions is unfeasible. To overcome this difficulty deriving a detailed model of the reactions inside the cell and aiming for mathematical simplicity, we take a similar approach as in previous studies \cite{mori2008wave,Camley17,Alonso18} and formulate a simple reaction-diffusion equation, where the non-linear reaction kinetics leading to bistability is modelled by a cubic polynomial in the variable $c(\boldsymbol{x},t)$. In addition, we introduce a term accounting for degradation of the biochemical component $c$. The equation reads 

\begin{equation}\label{eq1}
\frac{\partial (\phi c) }{\partial t} = \nabla \left(\phi D \nabla  c\right) + \phi [k_a\, c\,(1-c)(c-\delta(c)) - \rho\, c] + \,\phi\,(1-\phi)\,\xi(x,t),  
\end{equation}
where $k_a$ is the reaction rate, $\rho$ the degradation rate, and $D$ the diffusivity of the biochemical component. The last term on the right hand side introduces noise at the cell membrane, which allows us to account for the stochastic nature of the reaction-diffusion processes occurring within the cell. The noise intensity along with the reaction rate are key parameters in our model that allow the transition between different forms of cell motility. The stochastic variable $\xi(x,t)$  follows an Ornstein-Uhlembeck dynamics 
\begin{equation}\label{noise}
\frac{d \xi}{d t} = -k_{\eta}\, \xi + \eta\,,
\end{equation}
where $\eta$ is a Gaussian white noise with zero mean average $\langle \eta \rangle=0$ and a variance of $\langle \eta(x,t) \eta(x\sp{\prime},t\sp{\prime})\rangle=2\sigma^2\delta(x-x\sp{\prime})\delta(t-t\sp{\prime})$. 

The reaction diffusion equation aims at reproducing the pattern activity on the substrate-attached cell membrane observed in our experiments. A control of the size of the patterns is important since the patterned area rarely covers the entire cell membrane. Previous experiments with giant {\it D. discoideum} cells revealed that, after a critical size is reached, wave patterns tend to modify their shape rather than actually growing into larger areas \cite{Gerhardt4507}. Therefore, a dynamic control in the form of a global feedback on the parameter $\delta(c)$ is implemented. It affects the pattern dynamics and prevents the system to be covered completely by $c$ depending of the value of $C_0$. The control term reads
\begin{equation}\label{delta}
\delta(c) = \delta_0 + M\left( \int \phi\, c\, dA - C_0 \right)\,,
\end{equation}
where the parameter $C_0$ represents the average area covered by component $c$. The control mechanism shown in eq.~\eqref{delta} dynamically changes the value of the unstable fixed point of the system. This enforces that the amount of component $c$ inside the cell is constant on average. In our simulations, we will specifically show the effect of changing $C_0$ on the cell trajectories.

We integrated Eqs.~\eqref{pf}-\eqref{delta} on a square domain of 300$\times$300 pixels with periodic boundary conditions using standard finite differences. The pixel size is given by $\Delta x = \Delta y = 0.15$\,$\mu$m and the integration time step is $\Delta t=0.002$\,s. The values and definitions of the parameters of the model can be found in Table \ref{table1}. Cell trajectories and velocities were obtained by finding and tracking the center of mass of cells from the numerical simulations.

In the following sections, we will study and analyze different cell motility modes.
The transitions between these modes are obtained by varying three parameters: the noise intensity, the average membrane coverage with the activatory component $c$, and the activity rate of the biochemical field. In general, the bistable kinetics of $c$ will drive the formation of patches of high concentration of $c$ on a background of low $c$ concentration. The coherent effects of these patches on directed cell locomotion will be disturbed and interrupted by the impact of noise, which will favor nucleation events and the formation of new patches in other regions of the membrane.
Therefore, the dynamics of our model can be qualitatively understood as a competition between the coordinated effects of pattern formation and the randomizing impact of noise on cell locomotion.

%%%%%%%%%%%%%%%%%%%%%%%%%%%%%%%%%%%%%%%%%%%%%%%%%%%%%%%%%%%%%%%%%%
\begin{table}[]
	\caption{Parameter values for the numerical model.}\label{table1}
	\begin{tabular}{@{}*{5}{l}}
		Parameter  & Value   & Units   & Meaning  & Value Reference\\
		$D$        &   0.5  & $\mu m^2 /s$  &  Diffusion coefficient & \cite{Alonso18}\\
		$k_a$      &   2-5   & $s^{-1}$  &  Reaction rate   & \cite{Alonso18} \\
		$\rho$     &   0.02   & $s^{-1}$  &  Degradation rate	 	& \cite{Alonso18}\\
		$\sigma$   &   0.15    &  $s^{-2}$ &  Noise strength	      & \cite{Alonso18}\\
		$\tau$     &   2   & $pN s \mu m^{-2} $ & Membrane dynamics time-scale  & 1\cite{shao2010computational}, 2.62\cite{ziebert2011model}\\
		$\gamma$   &   2   & $pN$ & Surface tension	& 1\cite{shao2010computational}     	\\
		$\epsilon$ &   0.75   & $\mu m$ & Membrane thickness   & 1\cite{shao2010computational}\\
		$\beta$    &   22.22   & $pN \mu m^{-3} $ & Parameter for total area constraint & \cite{Alonso18} \\
		$A_0$ 	   &   113   & $\mu m^{2}$ &   Area of the cell & \cite{Alonso18}  \\
		$\delta_0$ & 0.5   & - &   Bistability critical parameter & \cite{Alonso18}\\
		$M$ 	   & 0.045 & $\mu m^{-2}$ &   Strength of the global feedback input & \cite{Alonso18}\\
		$k_{\eta}$ & 0.1   &  $s^{-1}$   &   Ornstein-Uhlembeck rate & \cite{Alonso18}\\
		$\alpha$   & 3 & $pN \mu m^{-1}$ & Active tension  & 0.5-3\cite{ziebert2011model}\\
		$Co$       & 28, 56, 84 & $\mu m^{2}$ & Maximum area coverage by $c$ & \cite{Alonso18} \\
		
	\end{tabular}
\end{table}
%%%%%%%%%%%%%%%%%%%%%%%%%%%%%%%%%%%%%%%%%%%%%%%%%%%%%%%%%%%%%%%%%%
\section{Results}
We performed a systematic study of the model described in the previous section. By modifying the values of the biochemical reaction rate $k_a$, the intracellular area covered by the concentration $c$, defined as $C_0$, and the noise strength $\sigma$, we obtained a diverse set of cell shapes, trajectories, and speeds. An overview of the studied cases is presented in Figure 1, where different cell shapes and average speeds are shown in the plane spanned by the parameters $\sigma$ and $C_0$. We identified four different types of motility with distinct shapes and trajectories: Amoeboid cells, characterized by a motion parallel to the elongation axis, fan-shaped cells that move perpendicular to the elongation axis, intermediate states that combine features of both amoeboid and fan-shaped types, and oscillatory cells, where the concentration $c$ is almost homogeneously distributed inside the cell with only small fluctuations at the border.

\begin{figure}[!]
	\centering
	\includegraphics[width=1.0\textwidth]{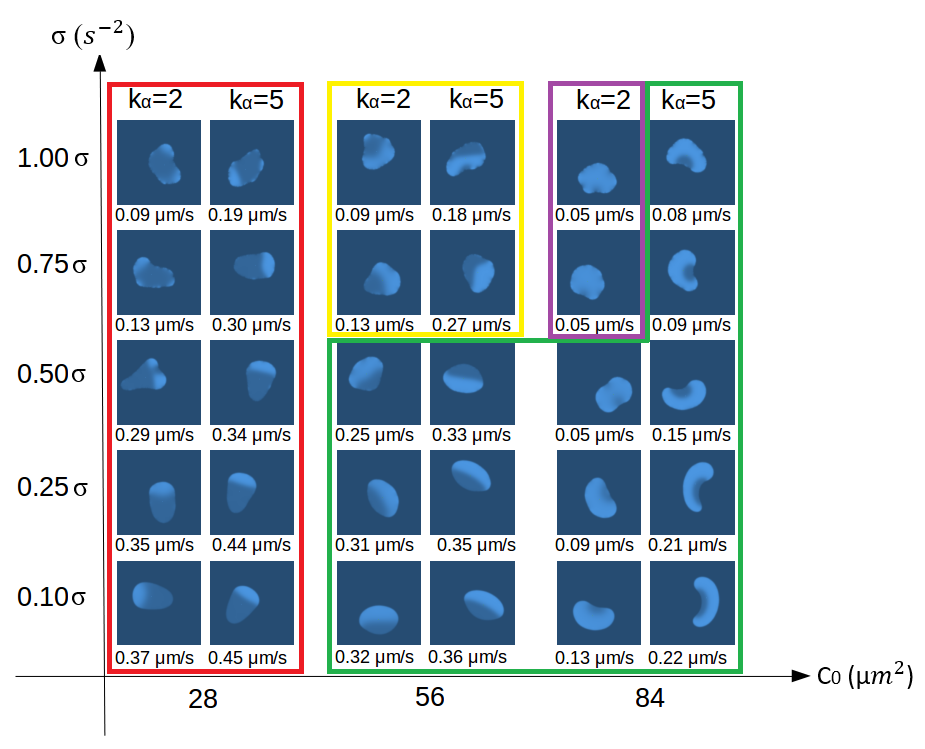}
	\caption{\label{Figura1} Phase diagram of the cell shapes obtained from the model and their respective average speeds for different values of the spatio-temporal noise, the reaction rate, and the maximum area coverage of the concentration $c$. The intracellular concentration $c$ is proportional to the intensity of sky blue color. Cell speed was computed as the sum of the velocities at each time step divided by the number of time steps. Colored boxes classify cells into four types: Amoeboid cells inside the red box, fan-shaped and similar cases inside the green box, intermittent cases at the transition from the amoeboid to the fan-shaped regime inside the yellow box, and oscillatory cells inside the purple box.}
	\label{fig:dfas}
\end{figure}

%In \autoref{fig:amosimexp} a more complete diagram is shown for all the intermediate cases for both reaction rate values used in the simulations. Here, the transition between the cases described in \autoref{fig:dfas} is seen in more detail.

\begin{figure}[!]
	\centering
	\includegraphics[width=1.0\textwidth]{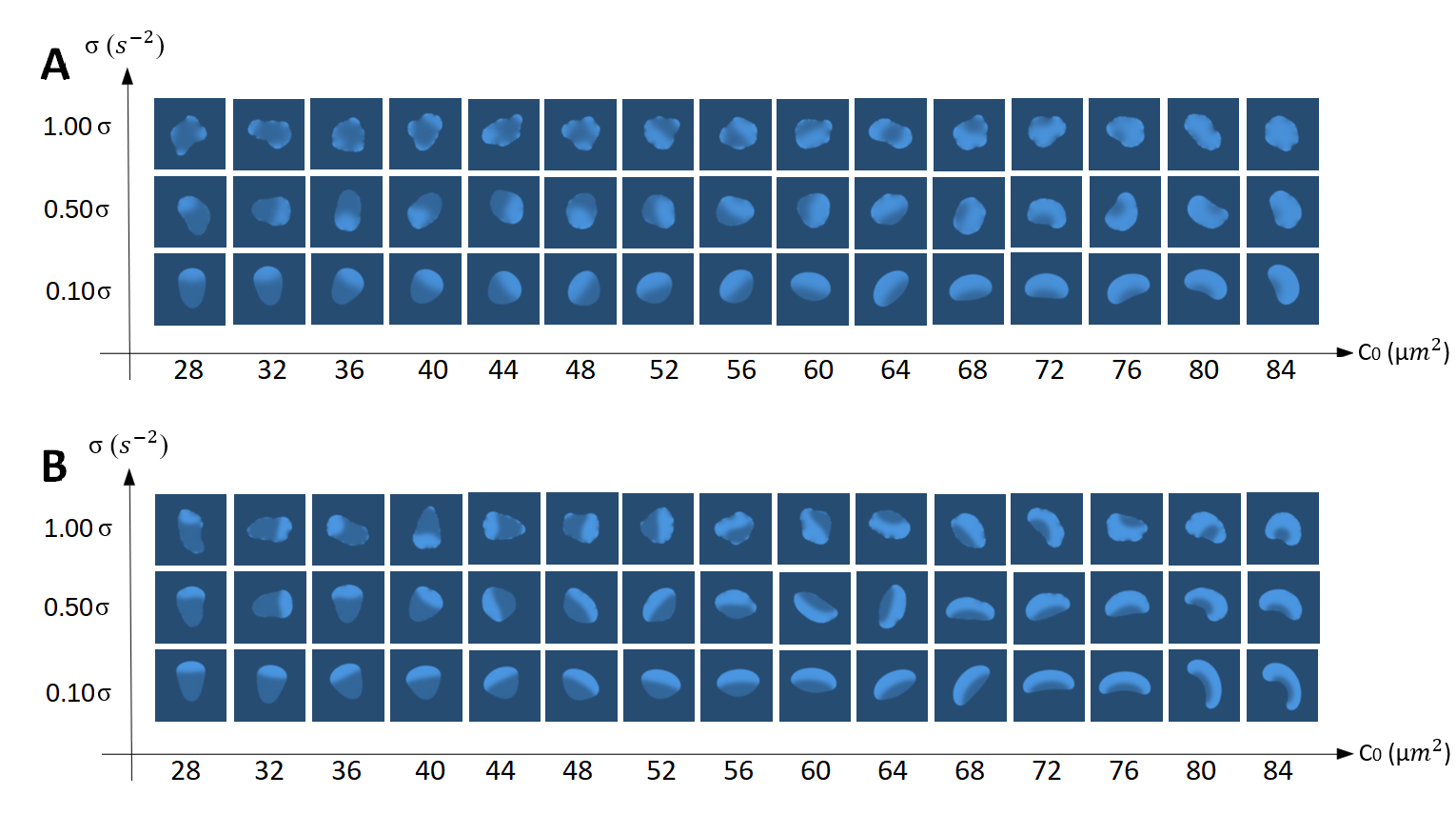}
	\caption{\label{Figura2} Phase diagrams of the cell shape obtained from the model simulations when varying area coverage and noise intensity. The computational cells are obtained for $k_a=2s^{-1}$ (A) and $k_a=5s^{-1}$ (B). The intracellular concentration $c$ is proportional to the intensity of sky blue color.}
	\label{fig:allcases}
\end{figure}

The transitions between the four cell types, summarized in \autoref{fig:dfas}, are presented in more detail in \autoref{fig:allcases}, where cell shapes are shown for $k_a=2s^{-1}$ and $k_a=5s^{-1}$ separately as a function of the parameter $C_0$, which is changed in smaller increments here. In both diagrams, the simulations in the left top panels show cell types that mimic the vegetative and starving states of {\it D. discoideum}, analyzed in more detail in \cite{Alonso18}. The right bottom panels in \autoref{fig:allcases} A and B correspond to stable fan-shaped and rotating fan-shaped cells, respectively, which are shapes related to particular mutations of {\it D. discoideum}  \cite{asano2004keratocyte,miao2017altering}. In between these two limits, different dynamical regimes have been obtained during the computational study of our model, some of which have been also observed in our experiments with {\it D. discoideum}, as we described below.

\subsection{Amoeboid motion.}

If we set $C_0=28$ $\mu m^2$, the biochemical component $c$ roughly occupies one quarter of the total cell area, fixed at $113$ $\mu m^2$ (corresponding to a circular cell with radius 6 $\mu m$). Under these conditions, the concentration $c$ accumulates in the front part of the cell reminiscent of the typical amoeboid shape of  {\it D. discoideum} cells. In this scenario, we observe different trajectories depending on the noise intensity. A large noise intensity translates into slow and random cell motion, whereas low intensity leads to faster and much more persistent motion. 
%While the simulations for $k_a=2s^{-1}$ show no clear resemblance to experiments, the simulations for $k_a=5s^{-1}$ show similar dynamics as the experiments \cite{Alonso18}.

\begin{figure}[t]
	\centering
	\includegraphics[width=1.0\textwidth]{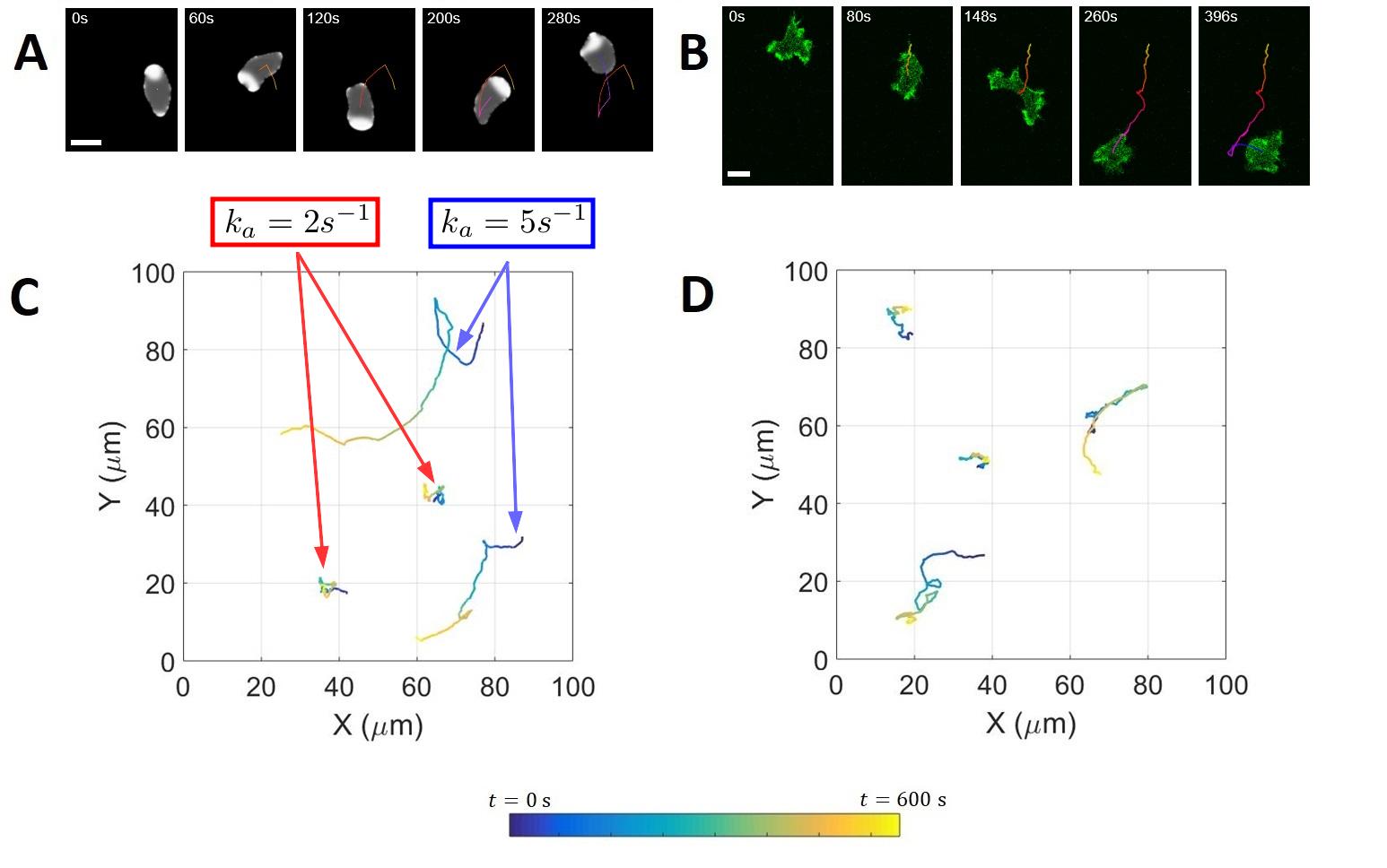}
	\caption{\label{fig:amosimexp}  Numerical and experimental results for amoeboid motion. (A) Sequence of several snapshots of a numerical simulation with $C_0=28\mu m^2$ and $k_a=5s^{-1}$. (B) Sequence of several snapshots of a vegetative DdB NF1 KO cell showing the amoeboid phenotype. (C) Example cell trajectories tracked from four simulations over 600 $s$ (two simulations with $k_a=2s^{-1}$ and two with $k_a=5s^{-1}$) It is important to mention that the two small trajectories correspond to $k_a=2s^{-1}$, while the two larger trajectories to $k_a=5s^{-1}$. Both cases also with $C_0=28\mu m^2$, (D) Examples of cell trajectories tracked from four vegetative DdB NF1 KO cells showing the amoeboid phenotype in experiments over 600 $s$.}
\end{figure}

A comparison between the dynamics of living  {\it D. discoideum} cells under amoeboid conditions and cell dynamics generated by our model is shown in \autoref{fig:amosimexp}.
The numerical simulations shown in \autoref{fig:amosimexp}A and C correspond to low values of $C_0$ and a noise intensity of 100$\%$ of the total given in Table 1. \autoref{fig:amosimexp}C, showing trajectories for $k_a=2s^{-1}$ and $k_a=5s^{-1}$, demonstrates that the parameter $k_a$ controls cell polarization \cite{Alonso18}. 
A value of $k_a=2s^{-1}$ induces a diffuse trajectory in a small area of space and a random appearance of protrusions formed by fluctuating amounts of concentration $c$ along the cell membrane. For $k_a=5s^{-1}$ the cell explores larger areas due to a continuous and more stable accumulation of $c$ in one region of the membrane that sets in the direction of motion (see \autoref{fig:amosimexp}A).
A representative sequence of snapshots of an experimental observation of amoeboid motion of a {\it D. discoideum} cell is shown in \autoref{fig:amosimexp}B, where the accumulation of actin is clearly appearing in the cell front. In addition, we present examples of individual experimentally recorded cell trajectories that cover the behavioral diversity of amoeboid {\it D. discoideum} cells \autoref{fig:amosimexp}D. 

%As we decrease the noise intensity in the numerical simulations the trajectories become more persistent, the cell shape more regular and a big concentration of $c$ appears in the edge of the cell, similar to a highly polarized state.  which is typically found in starving developing-cells. Some representative examples are shown in the last row of the first two columns in \autoref{fig:dfas}. 

\subsection{Intermediate dynamics.}

\begin{figure}[!]
	\centering
	\includegraphics[width=0.80\textwidth]{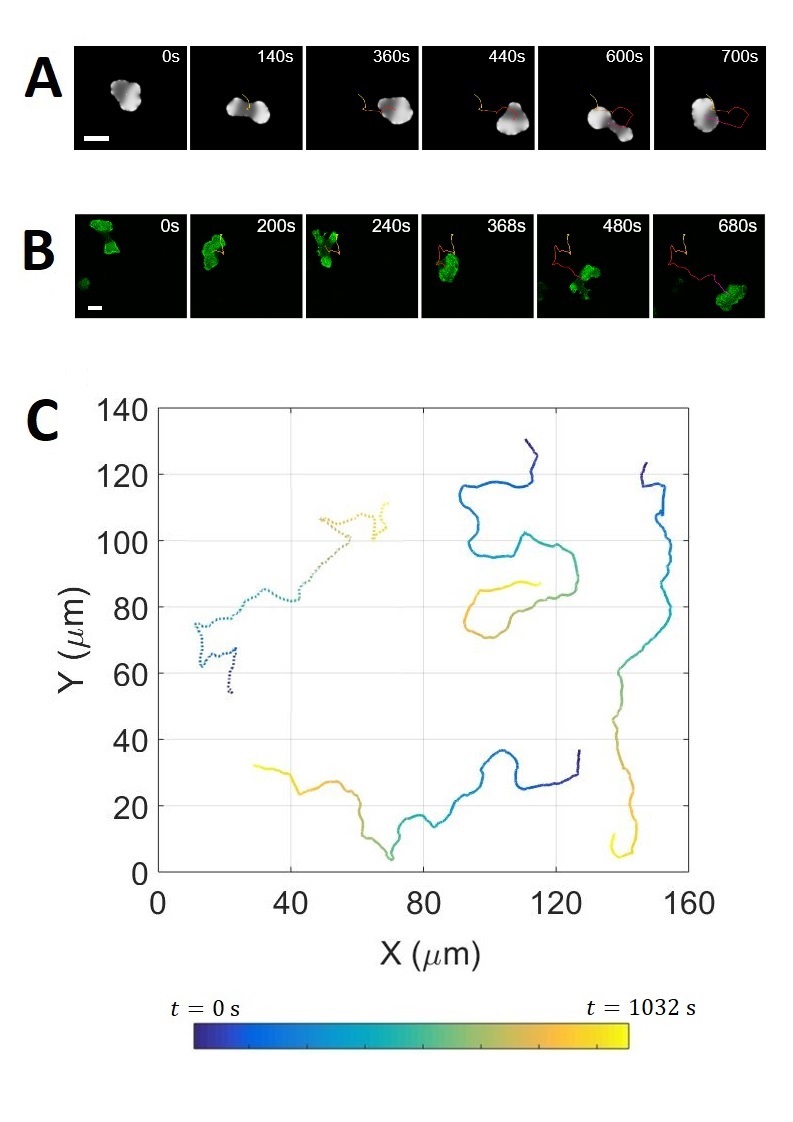} 
	\caption{\label{Figura1} Numerical and experimental results for the intermediate unstable case (A). Sequence of six snapshots of a numerical simulation with $C_0=56\mu m^2$, $k_a=5s^{-1}$. (B) Sequence of snapshots of a DdB NF1 KO cell starved for 4 hours prior to imaging, showing the intermediate phenotype. (C) Comparison between the trajectories of the center of mass of three numerical simulations (solid lines) and a trajectory of the center of mass of the DdB NF1 KO cell shown in B
		(dotted line). In all four cases, cells were tracked over $1032 s$.}
	\label{fig:unssimexp}
\end{figure}

By increasing the parameter $C_0$ to $56$ $\mu m^2$, which corresponds to half of the total area of the cell covered by the biochemical species $c$, and maintaining the noise intensity  between 75$\%$ and 100$\%$ of the total given in Table 1, we find a different motile behaviour in our model simulations. Under theses conditions, the results of the numerical simulations resemble the amoeboid shapes described in the previous section, however the repeated appearance of an additional large protrusion strongly modifies the trajectories of the simulated cells, see \autoref{fig:unssimexp}A. Initially, the amount of $c$ is concentrated in one region at the cell border, clearly defining a leading edge. From time to time, a part of the total amount of $c$ changes position at the cell border thus triggering an instability of the initial leading edge. This drives the formation of a new protrusion, where eventually most of the total amount of $c$ will accumulate and define a new cell front. In \autoref{fig:unssimexp}B, we show an example of similar intermittent behaviour that was observed in our experiments with {\it  D. discoideum} cells which frequently switch from amoeboid to fan-shaped motility and vice versa. \autoref{fig:unssimexp}C, where we present a comparison between an experimental trajectory and three trajectories obtained from numerical simulations, demonstrates how this dynamics generates trajectories with abrupt changes in direction.

%\begin{figure}[!]
% \centering
% \includegraphics[width=1.0\textwidth]{CasoUnstSimExp.jpg}
% \caption{\label{Figura1} Numerical and experimental results for the intermediate unstable case. (A) represents a sequence of three snapshots of the numerical evolution. (B).Snapshots obtained from experimental results. Three solid lines in (C) corresponds to trajectories obtained of numerical integration while dotted line corresponds to experimental results. For sumulation and experiment, cell trajectories were tracked over $1032 s$.}
% \end{figure}

By further increasing the parameter $C_0$ to $84$ $\mu m^2$, corresponding to 75\% of the total area of the cell covered by concentration $c$, another distinct behaviour is obtained in the numerical simulations, see the purple box in \autoref{fig:dfas}. Keeping $k_a=2s^{-1}$ and a noise intensity similar to the previous case, an oscillatory behaviour of the cell border is observed due to saturation of $c$ inside the cell. 
Noise-driven small displacements and a {\it circular} cell shape characterize this regime. It resembles previous experimental observations of a so-called pancake phenotype \cite{edwards2018insight}. 
%A noise driven small displacements of the cells and a  {\it pancake} shape of the cell characterizes this regime comparable with some previous experimental observations \cite{edwards2018insight}. 

% From %%%%Agregar o no caso oscilatorio 

\subsection{Fan-shaped motion.}

For the values of $C_0$ employed in the previous section but low noise intensity, the shape of the numerically obtained cells becomes more elongated  perpendicular to the direction of motion than parallel to the direction of motion of the cell. Moreover, their elongated shape is stable over time, and they move in a highly persistent fashion. Together these features characterize the so-called fan-shaped motion of {\it D. discoideum} cells~\cite{miao2017altering}.
The overall appearance and motion characteristic of fan-shaped cells share many similarities with keratocytes, even though the internal organization of the motility apparatus is clearly different.

% This case has the characteristic that is present in two different scenarios of our classification. 

%The first condition considered is for 
For $C_0=56\,\mu m^2$, corresponding to a concentration $c$ covering  half of the total cell area, and a noise intensity that is reduced to 50$\%$ or less of its maximal value,
%. Under such conditions 
a rounded elongated shape, reminiscent of a keratocytes, is observed in the numerical simulations, see for example the results in \autoref{fig:dfas}. The trajectories are straight and persistent for lower noise intensities and become more erratic for high noise levels.

\begin{figure}[t]
	\centering
	\includegraphics[width=1.0\textwidth]{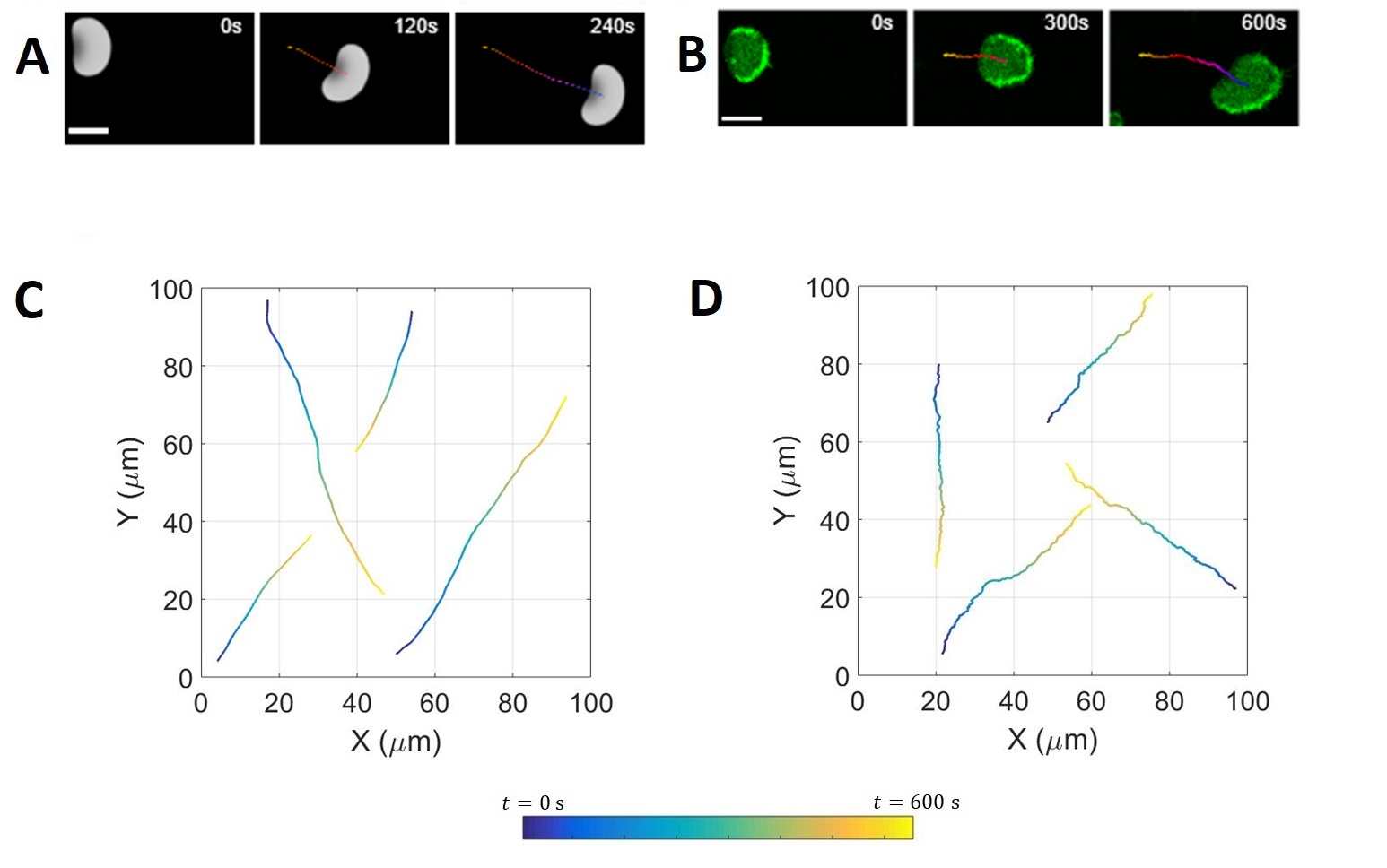}
	\caption{\label{Figura1} Numerical and experimental results for fan-shaped motility. (A) Sequence of three snapshots taken from a numerical simulation of the computed cell with parameter values $C_0=84\mu m^2$, $k_a=2s^{-1}$ and noise intensity set to $10$ percent. (B) Snapshots of a DdB NF1 KO cell starved for 4 hours prior to imaging, showing the fan-shaped phenotype. (C) Four examples of trajectories of the center of mass of numerically simulated fan-shaped cells tracked over $600$ $s$. (D) Four trajectories of the center of mass of DdB NF1 KO cells starved for 4 hours prior to imaging, showing the fan-shaped phenotype tracked over 600 $s$ in experiments.}
	\label{fig:kersimexp}
\end{figure}

By further increasing the covered area to $C_0=84\,\mu m^2$, we obtain similar fan-shaped cells. For $k_a=2s^{-1}$ the simulation produces rounded cell shapes that move at a reduced speed in a highly random fashion.
Comparing \autoref{fig:kersimexp}A and B reveals the qualitative similarities between fan-shaped cells obtained from simulations under these conditions and the experimentally observed dynamics of  fan-shaped {\it D. discoideum} cells. The model satisfactorily reproduces the experimental features of the cell motion. The four different realizations of trajectories, generated in numerical simulations and presented in \autoref{fig:kersimexp}C, display a similar persistent motion as the straight cell trajectories observed in experiments, see \autoref{fig:kersimexp}D. These trajectories resemble the trajectories of fan-shaped cells with $C_0=56\mu m^2$ and intermediate noise intensity as discussed previously. 
%
%
%
%The second scenario is related to coverage parameter which produces a transition between persistent and rotational motion.
%
Thus, depending on the value of the reaction rate and the noise intensity, we see remarkable differences between cell shapes and trajectories.

\subsection{Rotational trajectories of fan-shaped cells.}

\begin{figure}[t]
	\centering
	\includegraphics[width=1.0\textwidth]{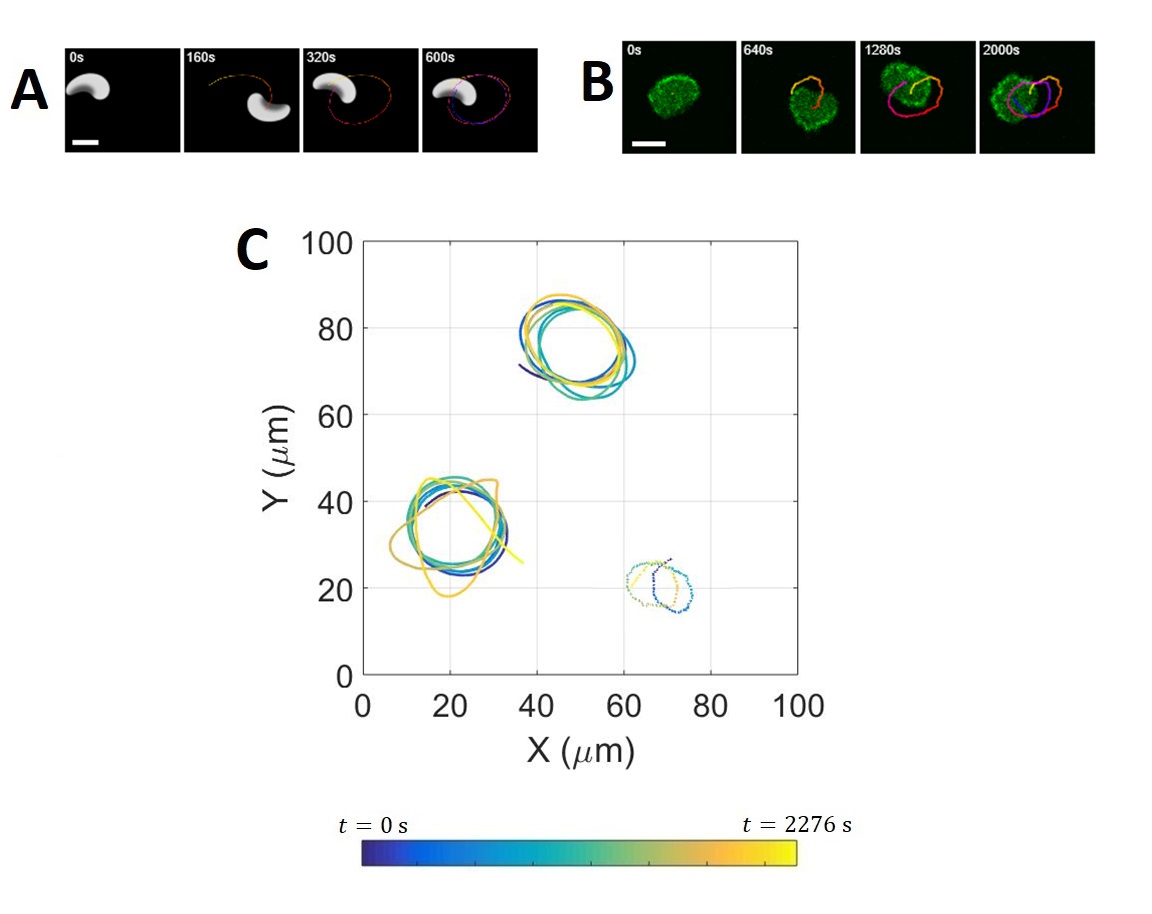}
	\caption{\label{Figura1} Numerical and experimental results for the rotational fan-shaped case. (A) Sequence of four snapshots obtained in the numerical simulations for $C_0=84\mu m^2$, $k_a=5s^{-1}$ and noise intensity set to $10$ percent. (B) Sequence of snapshots of a DdB NF1 KO cell starved for 5 hours prior to imaging, showing a fan-shaped cell with rotational movement. (C) Comparison of the trajectories of two simulations (solid lines) and one experimental realization (dotted line). The three lines  correspond to trajectories tracked over more than $2000 s$.}
	\label{fig:rotsimexp}
\end{figure}

Numerical simulations with $C_0=84\mu m^2$ and $k_a=5s^{-1}$ produce fan-shaped cells with a more elongated and curved shape. Depending on the noise intensity, different scenarios are obtained ranging from irregular shapes and trajectories at high levels of noise to regular shapes and circular trajectories for lower noise levels (see \autoref{fig:dfas}). Under these conditions the trajectories may also reveal rotational dynamics. In \autoref{fig:rotsimexp}A and B, we show examples of rotational dynamics observed in a simulation and in an experiment, respectively, finding a qualitative similarity between them. The corresponding trajectories are displayed in \autoref{fig:rotsimexp}C for comparison, along with a third trajectory of another simulation. Despite the differences in radius and frequency of rotation between simulations and experiment, the main characteristics of a periodic rotary motion are reproduced. Note that the concentration patterns inside the simulated cells resemble a half-moon shape, which is typical for fan-shaped cells with both straight and rotational trajectories.

\begin{figure}[!]
	\centering
	{ \includegraphics[width=1\textwidth,height=8cm,keepaspectratio]{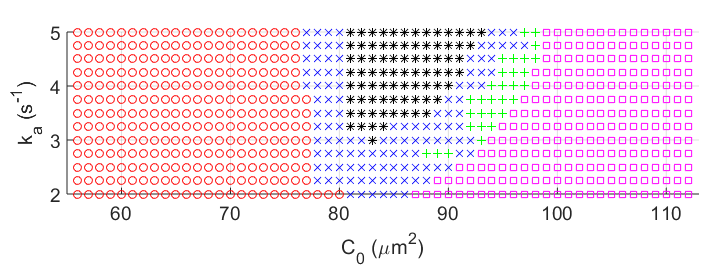}}
	\caption{\label{Figura2} Phase diagram representing the entire range of different dynamics obtained from the model at low noise intensity and as a function of the reaction rate $k_a$ versus the maximum area coverage $C_0$. Red circles represent trajectories that combine straight and curvature paths (fan-shaped cells). Blue crosses indicate irregular circling trajectories. Black dots show the region where regular circular trajectories are found. Green marks represent curved paths. Pink squares represent trajectories of almost stationary cells (pancakes). The noise intensity was set to 10\%. }
	\label{fig:dfln}
\end{figure}

%Once we see that straight trajectories resemble to fan shape of crawling keratocytes,  we showed that changing certain parameters this model reproduce circular trajectories 777a variability of cell shapes and in addition linear to circular trajectories. 
%Patterns of concentration $c$ inside the simulated cells with both straight and rotational trajectories resemble a half-moon shape, which is typical in fan-shape cells.
The transition from straight to rotational motion for different values of $C_0$ and $k_a$ is shown in a phase diagram in \autoref{fig:dfln}. The noise intensity was kept constant at 10$\%$ in all cases. 
For low values of $C_0$ the trajectories of the simulated cells are straight and only sometimes exhibit a slight curvature, depending on the realization and the parameter values.
With increasing parameter $C_0$, irregular trajectories are observed combining straight pieces with rapid rotations giving rise to a highly erratic motion. 
For values of $C_0$ between 80$\mu m^2$ and 90$\mu m^2$ and $k_a$ larger than 3$s^{-1}$, the simulations produce rotating cells as shown in \autoref{fig:rotsimexp}.
For even higher values of $C_0$, after a small region with curved trajectories, the cell surface is almost saturated with the concentration $c$, and due to the low noise intensity, no significant net motion is observed.

\begin{figure}[!]
	\centering
	{ \includegraphics[width=1\textwidth,height=5cm,keepaspectratio]{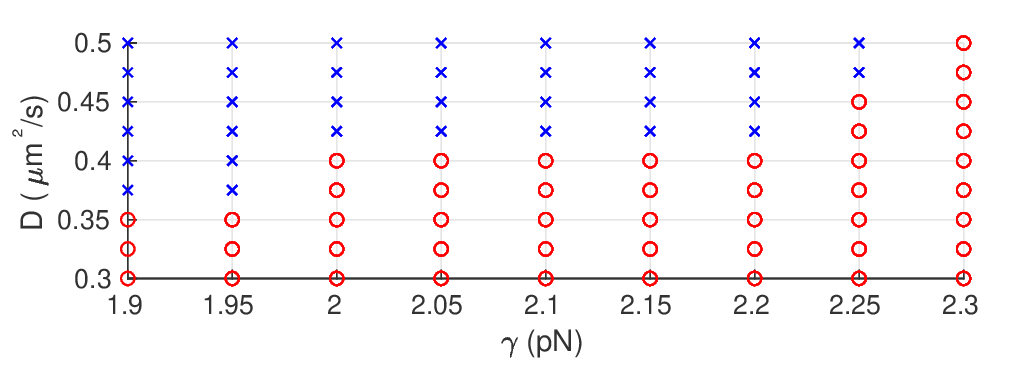}}
	\caption{\label{Figura2} Phase diagram in the plane spanned by the diffusion coefficient $(D)$ and the surface tension $(\gamma)$ of the numerical simulations of fan-shaped cells with straight and circular trajectories corresponding, respectively, to red circles and blue crosses. Parameter values are: reaction rate $k_a=5s^{-1}$,  coverage area $C_0=84\mu m^2$, and a low level of noise intensity (10\%).}
	\label{fig:dfDif}
\end{figure}

Thus, we have found that rotational trajectories arise for specific combination of the reaction rate and the area coverage, and that they are favoured by low values of the noise intensity. Finally, we also investigate if rotational trajectories can be induced by other factors. In \autoref{fig:dfDif}, we present a trajectory phase diagram spanned by the diffusion coefficient ($D$) and the surface tension ($\gamma$). The simulations indicate that rotational modes giving rise to circular trajectories are obtained by increasing the diffusion coefficient and by reducing the surface tension. This analysis agrees  with the results obtained with a similar model for keratocyte dynamics described in \cite{Camley17}.   

\subsection{Cell shapes and velocities in numerical simulations are comparable to typical experimental values.}

As we described in the previous sections, shapes and trajectories of the cells vary strongly from one case to another. Amoeboid cells produce fluctuating displacements, fan-shaped cells exhibit also persistent and rotational motion. Finally, an intermediate case between amoeboid and fan-shaped phenotypes produces motion with characteristic features of both cases. The great majority of shapes and dynamics have also been observed in experiments, and a good qualitative agreement between experimental and numerical results was found.

In this section we will perform a more quantitative comparison between the experimental and the numerical results, and we define an index which clearly differentiates amoeboid and fan-shaped cells and permits to characterize the transition between both cases. There are several indices that are commonly used in the studies of cell migration \cite{gorelik2014quantitative}, such as the directionality ratio, the mean square displacement (MSD), and the directional autocorrelation. Here we will compute the directionality ratio to characterize the transition between amoeboid and fan-shape.
It is defined as the distance between the starting point and the endpoint of the cell trajectory, divided by the length of the real trajectory. Mathematically is defined as follows:

\begin{equation}\label{DRa}
\ DR = \frac{\left |  \vec{X_N}-\vec{X_0}\right |}{\sum_{n=0}^{N-1} \left |  \vec{X_{n+1}}-\vec{X_n}\right |},
\end{equation}

where $X_0$ and $X_N$ are the initial and final positions, respectively.

This ratio is close to 1 for a straight trajectory and close to 0 for a highly curved trajectory.

\begin{figure}[!]
	\centering
	{ \includegraphics[width=1.0\textwidth]{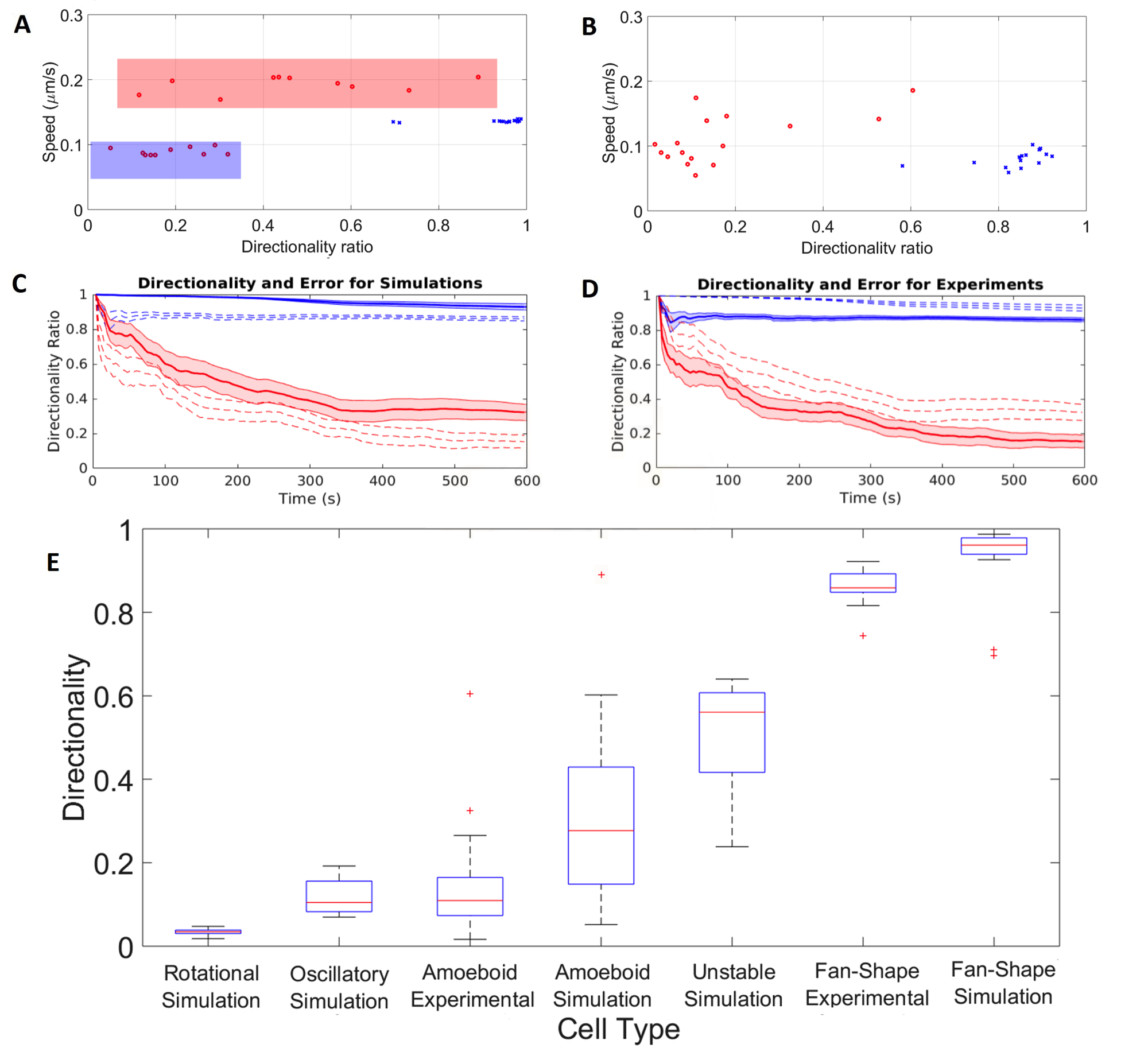}}
	\caption{\label{Figura2} Directionality ratio from simulations and experiments over 600$s$. (A) Cell speed versus directionality ratio obtained from numerical simulations of a sample of 20 amoeboid cells (red circles): 10 cells corresponding to $k_a=2$, $C_0=28\mu m^2$, and 100\% noise intensity (blue shaded area) and 10 corresponding to $k_a=5$, $C_0=28\mu m^2$, and 100\% noise intensity (red shaded area); and 15 fan-shaped cells (blue crosses). (B) Cell speed versus directionality ratio obtained from experiments with a sample of 16 amoeboid cells (red circles) and 15 fan shape cells (blue crosses) . (C) The mean value of the directionality ratio and standard error is plotted over time for amoeboid (red solid lines) and fan-shaped (blue solid lines) cells obtained from numerical simulations. Experiments curves from panel (D) are shown as dotted lines. (D) The mean value of the directionality ratio and standard error is plotted over time for amoeboid (red solid lines) and fan-shaped (blue solid lines) cells obtained from experiments. Simulation curves from panel (C) are shown as dotted lines.  (E) Box plot of the directionality ratio for the data of several simulations and experiments over $600s$.}
	\label{fig:Dir}
\end{figure}

%%parte Direccionalidad y vel
%There exist some indices that are very common to study the process of cell migration \cite{gorelik2014quantitative}, such as directionality ratio, mean square displacement (MSD) and direction autocorrelation. Being the first one a method easy to understand and also to compute.

%We calculated directionality ratio for the in vivo and in silico cases for both amoeboid shape and keratocyte-fan shape .

%We calculated directionality ratio for amoeboid shape and keratocyte-fan shape.
%DUDA EN ESTE PARRAFO%
In the following results DR was computed for every $\Delta t$ using the positions of the cell trajectories.
First, we compare the directionality ratio for a set of simulations corresponding to vegetative amoeboid cells ($k_a=2s^{-1}$), which produce low values of the directionality  and low velocities because of their random dynamics. Here, the relation between random motion and velocity can be interpreted as follows. A low noise intensity or large activity rate will lead to less nucleation events of new patches of $c$ at the membrane, favoring stable movement in one particular direction which will result in higher velocities. On the contrary, a high noise or small activity rate will generate more nucleation events at different positions of the membrane, that will compete with each other, thus pushing the cell membrane in different directions, resulting in lower instantaneous velocities.

Second, a set of simulations corresponding to starvation-developed amoeboid cells ($k_a=5s^{-1}$) is considered, which produce intermediate values of the directionality ratio because of their more persistent motion. Finally, we analyze a set of fan-shaped cells, which produce large directionality ratios because of their highly persistent movement. All these results are presented in \autoref{fig:Dir}A, where we see that the three types of cells are located in different regions of the parameter space.

The numerical results of the directionality ratio are in good agreement with the experimental measurements of this quantity, see \autoref{fig:Dir}B, where the experimental data is plotted for a similar amount of cells. 
Fan-shaped cells in both cases have large directionality ratios due to their persistent motion. The set of amoeboid cells naturally divides into two subsets: one with low directionality ratios and low speeds and the other one with larger directionality ratios (although lower than in the fan-shaped cases) and speeds spreading over a wide range. In the numerical simulations, we have produced equivalent subsets by changing the parameter $k_a$. % in the simulations. 

Further similarities between the numerical results and the experimental realizations are found when comparing panels C and D in \autoref{fig:Dir}. Here, the average of the directionality ratio over time with their respective errors for the  amoeboid and the fan-shaped cells are shown for the numerical simulations in \autoref{fig:Dir}C  and the experimental recordings in \autoref{fig:Dir}D. It is important to mention that for a better comparison we included experiments curves from panel D in C and curves from panel C in D, in both cases as dotted lines and also with their respective errors. The good agreement between the two cases is remarkable although  a systematic slight decrease of the directionality ratio appears in the experimental case, which may be related to the more noisy dynamics of the cell outline in the experimental measurements.

In \autoref{fig:Dir}E we display the directionality ratio for several cases obtained in the  simulations and for the two experimental cases discussed above. We compare the directionality ratios in a box plot representation of the different cases. First, we observe that the circular motion of fan-shaped cells gives rise to the smallest value of the directionality ratio as expected, because the motion is confined to a small region of space. The second observation is that intermediate cases, discussed in the previous sections, give rise also to intermediate values of the directionality ratio.  
The largest directionality ratios are observed for the stable fan-shaped cells. Note that we did not take into account other experimental cases because of the low number of recordings available in some of the experiments.

Finally, due to the cell shape diversity in experiments and simulations we have also analyzed the shape in a quantitative way. Computing quantities such as aspect ratio, ellipticity, or circularity are commonly computed when comparing cell shapes. There exist some earlier works, where cell morphology has been analyzed \cite{collenburg2017activity, lustig2019noninvasive, lee2014automated, frank2016frequent}. In this work, we focus on the circularity measure, which quantifies how closely the shape of a marked region approaches that of a circle. Circularity can be valued between 0 and 1, where 1 represents the value of a perfect circle. Mathematically the
circularity is defined as follows: $CR=\frac{4\pi A}{P^2}$, where A is the area and P is the perimeter of the cell. Along the trajectories of the cells, we calculate in each frame the value of the circularity for amoeboid and fan-shaped cells with the images obtained from simulations and experiments.

\begin{figure}[!]
	\centering
	{ \includegraphics[width=1.0\textwidth]{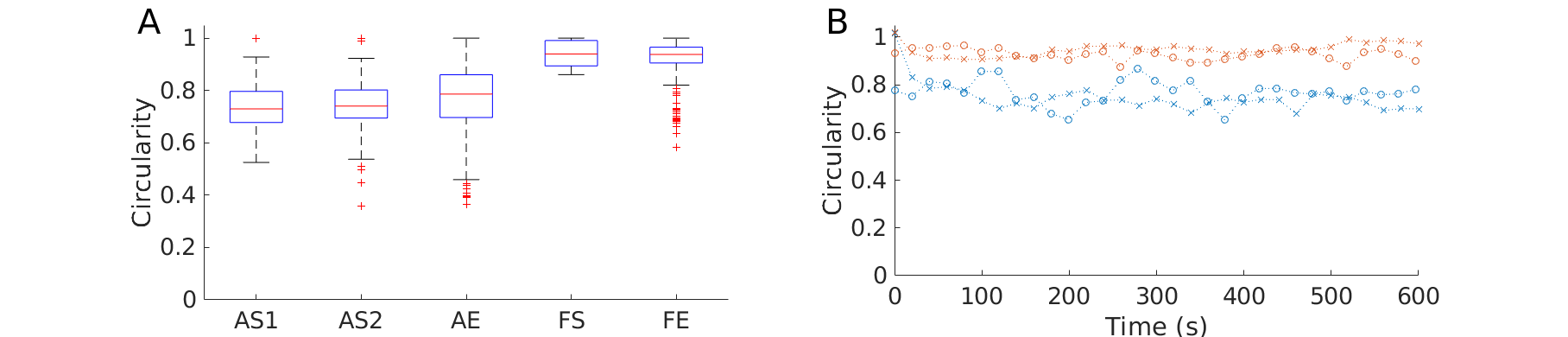}}
	\caption{\label{Figura2} (A) Circularity box plot representation of amoeboid and fan-shaped cells from simulations and experiments. Values of circularity for the results of 11 simulations and 10 experiments are shown: amoeboid simulation $k_a=2s^{-1}$ (AS1, n=3), amoeboid simulation $k_a=5s^{-1}$ (AS2, n=3), and fan-shaped simulations (FS, n=5), amoeboid experiments (AE, n=5), and fan-shaped experiments (FE, n=5). (B) Represents the mean value of circularity over time for the studied cases. Amoeboid and fan-shape simulations are represented in blue and red crosses, repectively. Red and blue circles corresponds to experiments of amoeboid and fan-shape cell, respectively.  The time for all the trajectories is 600$s$.}
	\label{fig:Cir}
\end{figure}

In \autoref{fig:Cir} we show a box plot representing the circularity ratio for the cases mentioned above. Here we see the tendency of fan-shape cells to oscillate around values close to 1 due to their rounder shape. On the other hand, amoeboid cells present a larger variation in the value of their circularity parameter because of their irregular fluctuating shape. From the results we also notice a good agreement between simulations and experiments for both scenarios. Just a small difference is marked in the fan-shaped case, where we obtained a larger amount of outliers in the analysis of the experimental data.

\begin{figure}[t]
	\centering
	\includegraphics[width=1.0\textwidth]{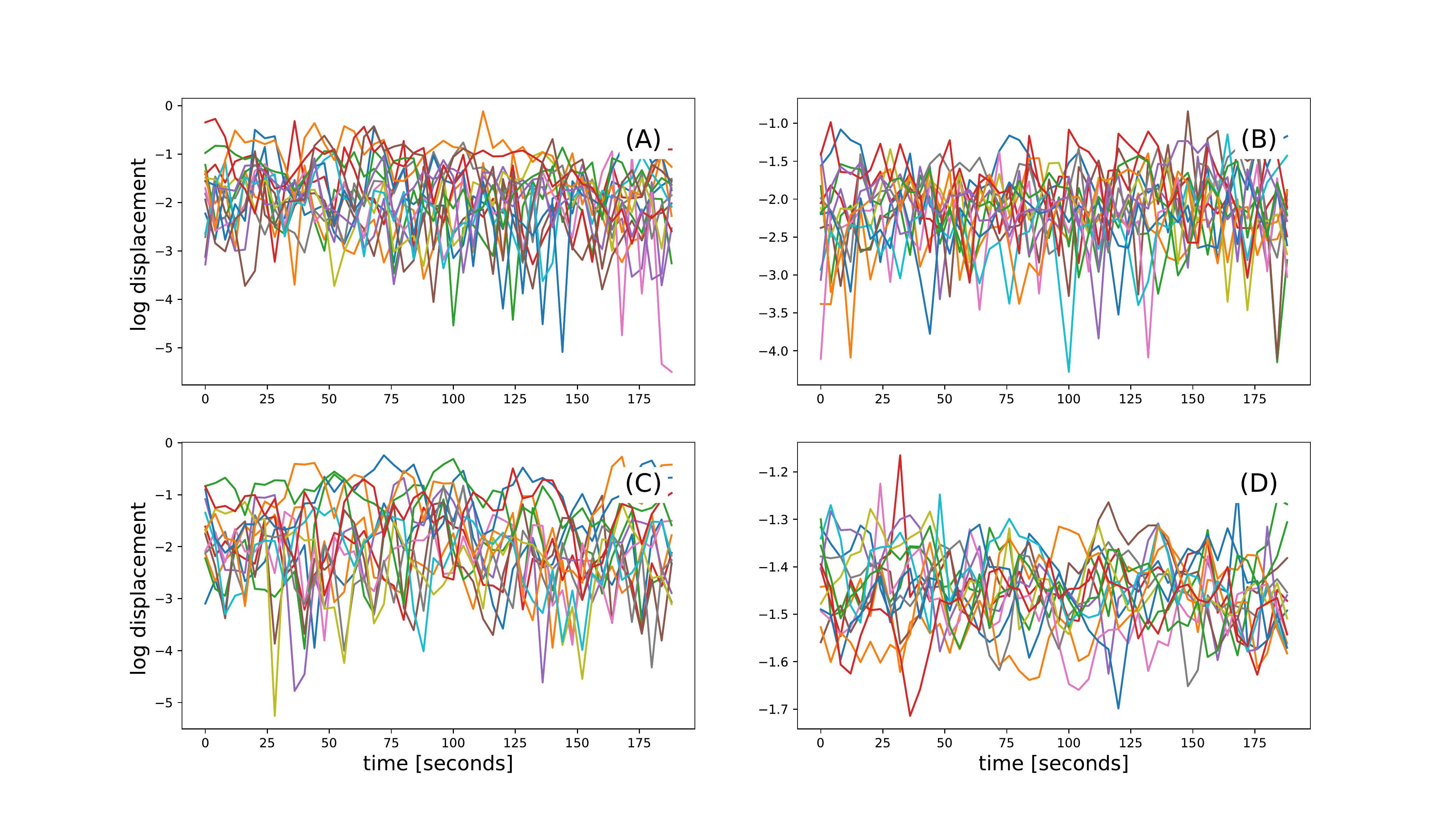}
	\caption{Time series of the log amplitudes of displacement vectors. (A)~Amoeboid and (B)~fan-shaped cells recorded in experiments. (C)~Amoeboid and (D)~fan-shaped cells produced by numerical model simulations. We consider 15 different trajectories for each type of cell.
		\label{fig:timeseries_amplitude}}
\end{figure}

\subsection{Correlation analysis of cell trajectories}

\begin{figure}[t]
	\centering
	\includegraphics[width=1.0\textwidth]{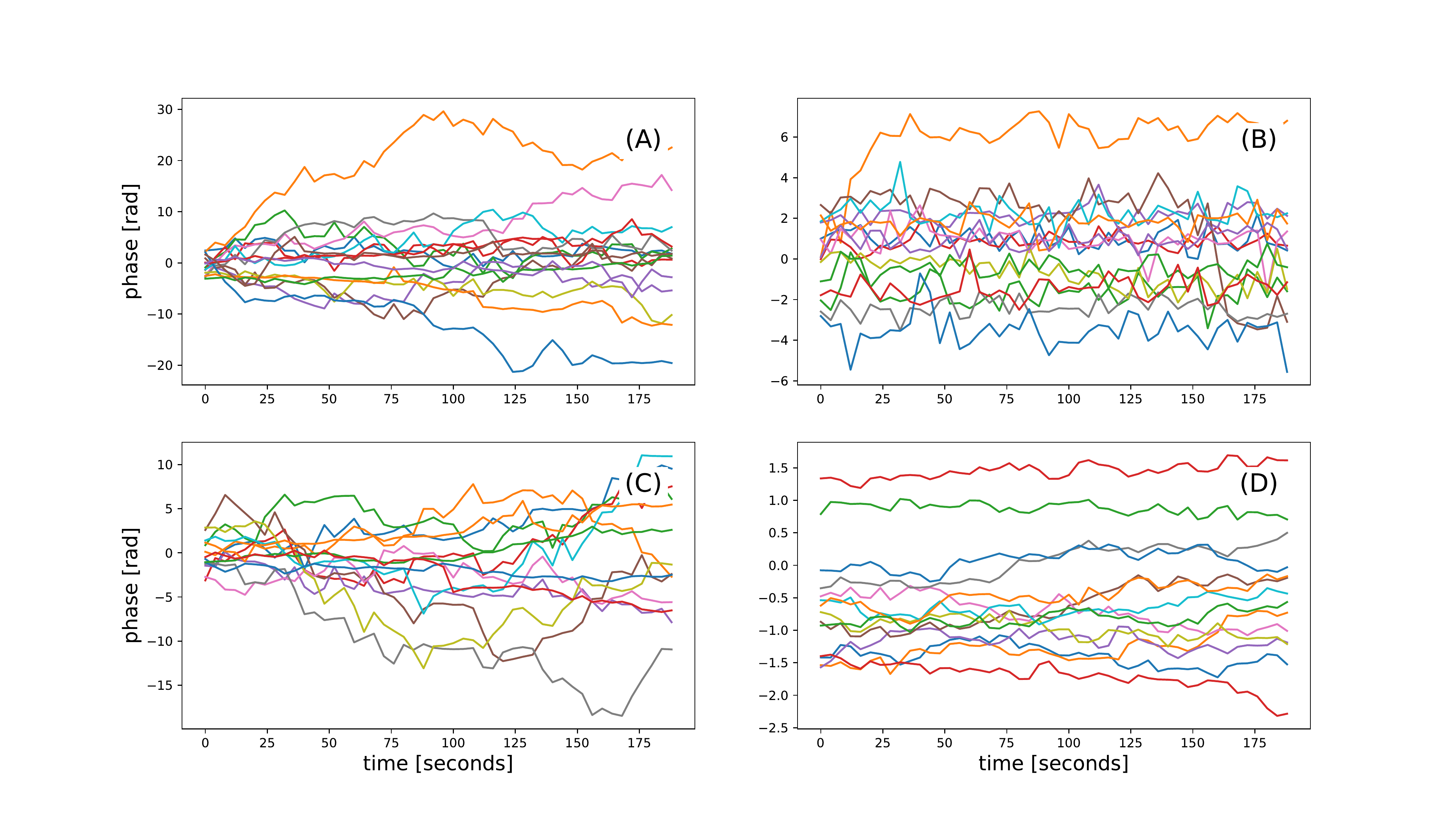}
	\caption{Time series of the phases of displacement vectors. (A)~Amoeboid and (B)~fan-shaped cells recorded in experiments. (C)~Amoeboid and (D)~fan-shaped cells produced by numerical model simulations. We consider 15 different trajectories for each type of cell.
		\label{fig:timeseries_phase}}
\end{figure}

To compare the cell trajectories obtained from model simulations and experimental recordings in more detail, we analyzed the correlation structure of the trajectories of both amoeboid and fan-shaped cells.
For this we chose a representation in polar coordinates, so that all the displacement vectors that connect the adjacent data points of a trajectory are represented by an amplitude (absolute value of the displacement) and a phase (angle with respect to the laboratory frame).
The time series of the log amplitudes and the phases are displayed respectively in \autoref{fig:timeseries_amplitude} and~\autoref{fig:timeseries_phase}.
In the case of the log amplitudes, the time series are stationary and fluctuate around a constant mean value.
Overall, the magnitudes and time scales of fluctuations are comparable between simulations and experiments and also between amoeboid and fan-shaped cases.
Only in the case of the simulated fan-shaped cells the magnitude of fluctuations is smaller.
In contrast to the amplitude, the time series of the phases is not stationary.
They furthermore reflect that the amoeboid cases reorient much more rapidly compared to the fan-shaped cases, i.e. in the given time interval, the phase drifts over a much larger range for the amoeboid cells than for the fan-shaped cells.
This difference is particularly pronounced for the simulated fan-shaped cells, where the phase remains almost constant over the entire measurement time, see~\autoref{fig:timeseries_phase}D. 

From the time series, we computed the corresponding autocovariances. For a real valued scalar time series $X_i$, $i=0,\dots, n-1$  of length $n$ we take the following estimator of the autocovariance ($|k| < n-1$)

\begin{equation}\label{ACV}
\hat\gamma(k) = \frac{1}{n-|k|} \sum_{i=0}^{n-|k|-1} ( X_{i+|k|} - \mu)(X_i - \mu),\quad \mu = \frac{1}{n} \sum_{i=0}^{n-1} X_i
\end{equation}

\begin{figure}[t]
	\centering
	\includegraphics[width=1.0\textwidth]{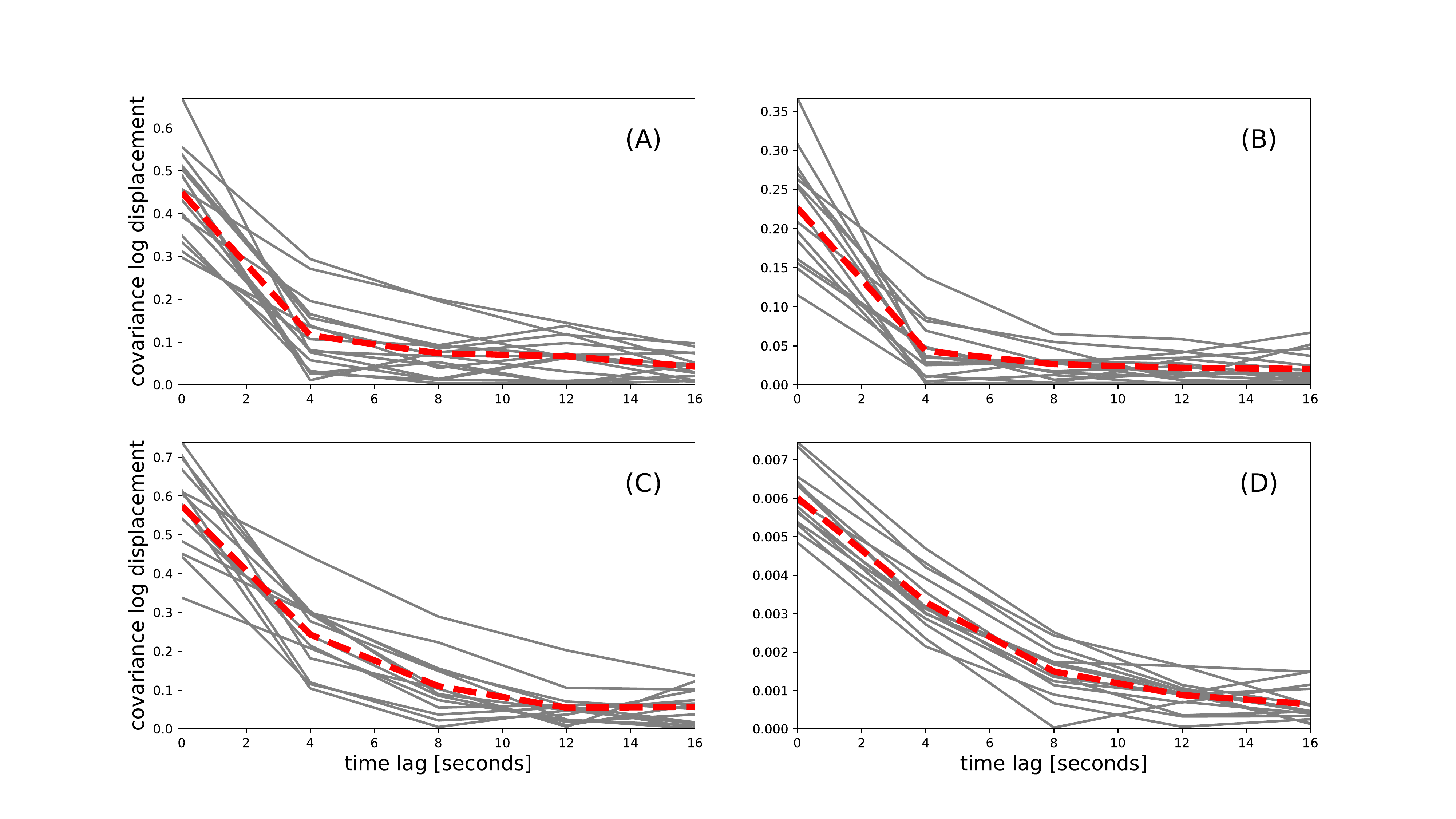}
	\caption{Autocovariance calculated from the time series of the log amplitudes of displacement vectors shown in~\autoref{fig:timeseries_amplitude}. (A)~Amoeboid and (B)~fan-shaped cells recorded in experiments. (C)~Amoeboid and (D)~fan-shaped cells produced by numerical model simulations. We consider 15 different trajectories for each type of cell. Dashed red thick lines correspond to the mean Autocovariances. 
		\label{fig:covariance_amplitude}}
\end{figure}

In~\autoref{fig:covariance_amplitude}, the autocovariance of the log amplitude is shown for both experimental and model trajectories of amoeboid and fan-shaped cells.
For the log amplitude of the model trajectories, we observe average correlation times that are slightly larger than for the experimental trajectories; they differ by a factor of approximately two. 
Nevertheless, in all cases the correlation time is rather short (of the order of seconds).
In particular, no significant difference is observed between amoeboid and fan-shaped cases.
Note, however, that the variances differ between amoeboid and fan-shaped cases, which is particularly pronounced in the case of the model trajectories.

\begin{figure}[t]
	\centering
	\includegraphics[width=1.0\textwidth]{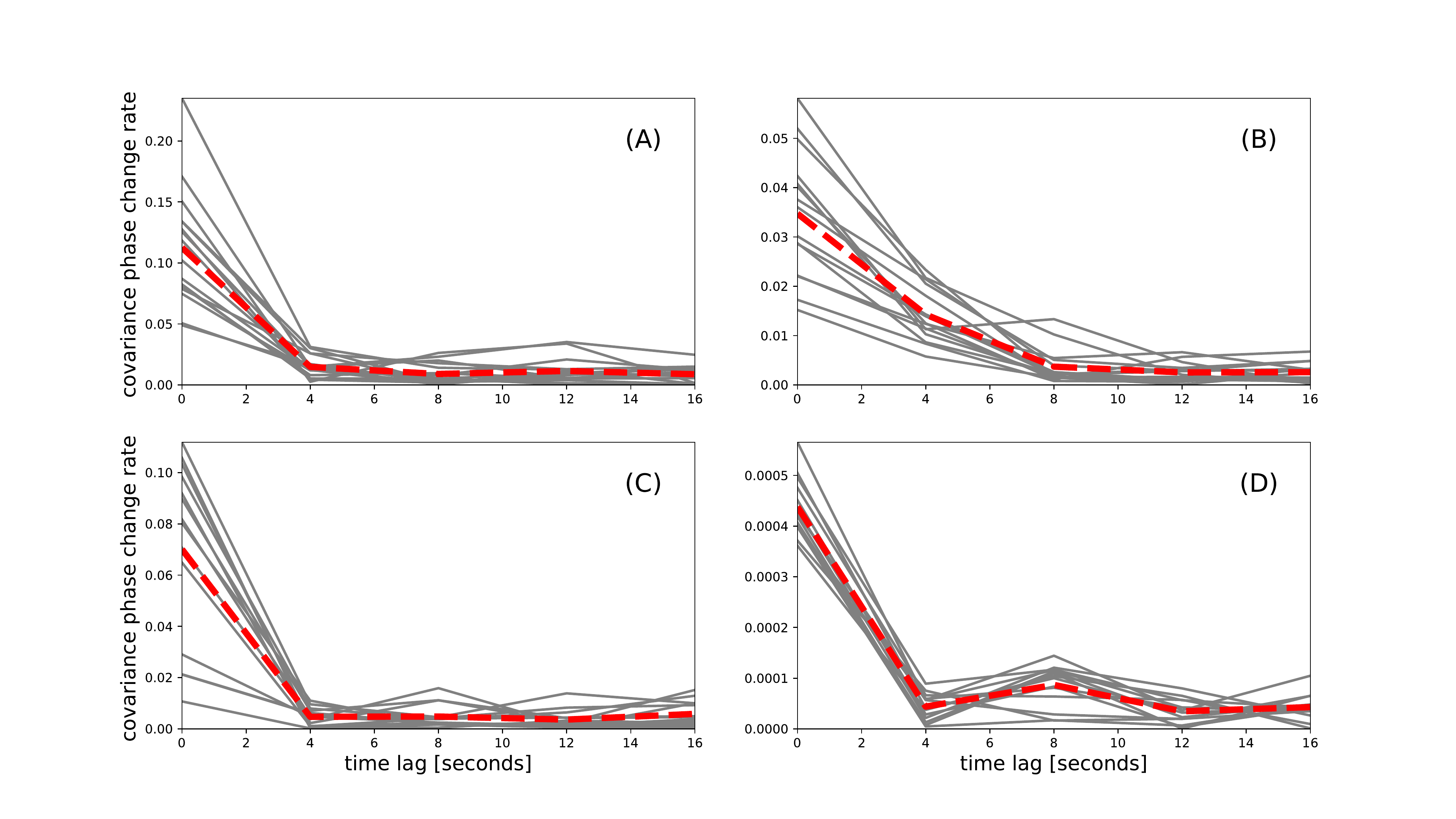}
	\caption{Autocovariance calculated from the time series of the phase change rate of displacement vectors. (A)~Amoeboid and (B)~fan-shaped cells recorded in experiments. (C)~Amoeboid and (D)~fan-shaped cells produced by numerical model simulations. We consider 15 different trajectories for each type of cell. Dashed red thick lines correspond to the mean Autocovariances.
		\label{fig:covariance_phase}}
\end{figure}

In the case of the phase, we calculated the autocovariance based on the time series of the rate of phase change between adjacent displacement vectors (the time series of the rate of phase change is stationary, while the phase time series is not).
In~\autoref{fig:covariance_phase}, the autocovariance of the rate of phase change is shown for all cases.
While in the amoeboid cases and for the simulated fan-shaped cells correlations decay to zero within 4~sec, the fan-shaped cells in the experiment show markedly larger correlation times, see~\autoref{fig:covariance_phase}B.
On the other hand, the variances are comparable in all cases except for the simulated fan-shaped cells, where the variance is by a factor of 100 smaller than in the other cases, see~\autoref{fig:covariance_phase}D.

From our correlation analysis, we thus conclude that the cell trajectories produced by our model simulations correctly capture the main correlation structure of the experimental trajectories. Only in the case of the fan-shaped cells, qualitative differences can be observed.
In the experimental case, an increased correlation time in the change rate of the phase leads to smooth and persistent trajectories.
The model trajectories, in contrast, do not show increased correlations in the phase change rate.
Here, smooth and straight trajectories originate from a strongly decreased variance of the phase change rate.

\section{Discussion}
%\subsection{Comparison with previous models in the literature}
We have studied a mathematical model that consist of biochemical dynamics in the form of a bistable reaction-diffusion equation including a noise description based on an Ornstein-Uhlenbeck process. The biochemical dynamics is coupled to a phase field to account for the deformable cell border. 
The model was previously introduced in \cite{Alonso18} to characterize the dynamics of vegetative and starvation-developed amoeboid {\it D. discoideum} cells. In that case, we found good agreement between the cell shape evolution, the intracelullar patterns, and the center of mass movement. 

Here, we have now systematically explored the entire relevant range of the parameter space and qualitatively reproduced different motility regimes observed in {\it D. discoideum} cells. %
The phase diagram of the model was explored by changing parameters such as the noise strength, the coverage area, and the rate responsible for cell polarization, giving rise to a series of different  motility scenarios as  presented above. The numerical results reproduced the dynamics of {\it D. discoideum} cells, which were difficult to catalog for the experimentalists. In general, the comparison of trajectories, cell shapes, and cell speeds between numerical simulations of the model and experimental data showed good agreement. 
We also performed a more quantitative comparison, based on the correlation structure of the amplitude and phase of the displacement vectors. 
Also in this case, good agreement was found and only minor differences in the case of the fan-shaped cells revealed that the model imposed smoothness of the cell trajectories of fan-shaped cells by reducing the variance of the phase changes rate, while the experimental data shows an increased correlation time instead.
In future studies, longer model trajectories will be generated to explore anomalous behavior in the diffusive properties that has been recently reported also for {\it D. discoideum} cells~\cite{makarava_quantifying_2014,cherstvy_non-gaussianity_2018}.
We should distinguish between the characteristic times obtained for example from the autocorrelation functions and the persistence time, which refer to different concepts. 
In contrast with the calculations using autocorrelation functions, the persistence times are typically calculated from mean square displacements (MSD) \cite{takagi2008functional,hiraiwa2014relevance}.

%By describing the displacement time series as an AR model of order one we were able to extract the model coefficients from the trajectories of amoeboid and fan shaped cases to compare between experiments and simulations. AR simulations with the extracted parameters yielded new trajectories in good agreement with the original ones.
% One could construct a link between the phase field and AR approximation by for example getting the mean of the AR parameters from a certain amount of experimental tracks then try to look for some phase field model simulation parameters (reaction rate, noise intensity and area coverage of $c$) of the whole map that reproduce the behaviour of the experimental tracks. Finally extract the AR parameters to compare with the ones obtained from the experiments and see how close are one to each other.%

%One could construct a link between the parameters of the phase field model and the parameters of the AR model and try to define the levels of $C_0$, noise intenisty and reaction rate $k_a$ that looks more similar .

The behavior of the model critically depends on the choice of the model parameters. Together with noise, realistic dynamics of intracellular patterns and cell shape changes are produced,  when correct characteristic temporal scales are used. The parameter $C_0$ corresponds to the area covered by the biochemical component $c$. It takes into account membrane deformations due to local accumulation of the biochemical component and, together with the reaction rate, reproduce variations in cell speed and persistence of motion.

Our model shows similarities to the bistable reaction-diffusion model coupled to a phase field described in~\cite{Camley17}, where the motion of keratocytes is investigated.
In particular, a systematic study of the transition between straight and circular motion of cells moving in a keratocyte-like fashion, equivalent to our fan-shaped cells, was done: a high tension tends to stabilize the cell motion to a straight trajectory, whereas large diffusion coefficients or small velocities tend to push the cell towards rotation~\cite{Camley17}. 
In our study presented here, we have investigated the connection of keratocyte-like behavior and amoeboid motility, apparently associated to different mechanisms and cell types.
We studied the transition between the persistent fan-shaped phenotype (keratocyte-like) and the amoeboid case, also showing intermediate dynamics and directly compared our model to experimental data obtained from recordings of {\it D. discoideum}.
With respect to the transition between straight and circularly moving fan-shaped cells our results shown in~\autoref{fig:dfDif} are consistent with the earlier predictions~\cite{Camley17}.
Transitions between amoeboid and fan-shaped motility modes have also been described recently in a wave-generating two-component reaction-diffusion model by Cao~\textit{et al.}~\cite{cao2019plasticity} that specifically emphasizes the role of cell deformation mechanics.
While our results are compatible with the findings of Cao~\textit{et al.}, our model clearly shows that the richness of different motility modes does not require intracellular traveling waves but can be already observed for intracellular kinetics that relies on a single dynamical variable only.

In contrast to these simple modeling approaches based on generic reaction-diffusion systems, there are also more complex descriptions following a different biophysical approach including more detailed biochemical reactions and mechanical forces.
For example keratocyte motion has been extensively studied in~\cite{maree2006polarization} combining biochemical and mechanical aspects to model how epidermal fish keratocytes form a leading edge, polarize, and maintain their shape and polarity. 
There are also more complex models where the transitions between straight and circular trajectories have been studied, see for example~\cite{Nickaeen17}, where a minimal mechanical model is presented consisting of two equations, one for the force balance of the actin network and a second one consisting of a reaction diffusion equation that describes the concentration of myosin, demonstrating that transitions occurs for small values of the Peclet number.

On the other hand, there are reductionist approaches to keratocyte motion, see for example~\cite{Camley17} and~\cite{ziebert2011model}, which display similar levels of complexity as our model.
Both types of descriptions contribute to a better understanding of the experimentally observed dynamics and can be readily extended in different directions.
For example, we are currently working on the implementation of more complex biochemical models into the phase field description. 
In particular, we can extend the model to more closely recover the detailed dynamics of certain intracellular reactions, such as, for example, the phosphorylation of PIP$_2$ to PIP$_3$ or the dynamics of the associated kinases and phosphotases that affect cell polarization, membrane deformation, and pseudopod formation~\cite{van2017coupled,fukushima2019excitable}.
Furthermore, the phase field framework will also allow us to implement cell-cell interactions~\cite{lober_collisions_2015}, the behavior of cells under confined stimuli~\cite{gerhardt_signaling_2014}, in enclosed environments~\cite{nagel_geometry-driven_2014,winkler_confinement_2019}, in the presence of external chemical gradients~\cite{najem2013phase}, and also in three dimensions~\cite{cao_minimal_2019}.

Here, we have restricted ourselves to a comparison of our model to {\it D. discoideum} cells.
However, we found that our model is also able to describe more diverse situations observed, for example, in keratocytes, where close to the transition to circular motion, bipedial motion was observed that relies on local alternation of cell displacements during persistent motion~\cite{Barnhart10}.

%In \cite{Nickaeen17} a minimal mechanical model is presented consisting in two equations, one for the force balance of the actin network and a reaction diffusion equation describing the concentration of myosin. The turning happens when Peclet number is small (i.e., small velocity scales) which is in agreement with the model results by Camely et al \cite{Camley17}.% Rotation for sufficiently small $\alpha$ (check what that means). They also measure the aspect ratio of the cell and find agreement with experimental data (can we do this?). 

\section{Conclusions}
In summary, we have studied a model based on a bistable reaction-diffusion equation with Ornstein-Uhlenbeck noise for the intracellular biochemistry, coupled to a dynamical phase field to describe the cell membrane dynamics.
The results obtained from the numerical integration of the model show that essential features of amoeboid and fan-shaped motion observed in experiments of motile {\it D. discoideum} cells are reproduced by our model.
We found close qualitative agreement between the numerical simulations and the experiments and, in some cases, motility measures such as the directionality ratio even showed quantitative agreement.
%Approximating the cell displacements as an AR(1) process 
The study of the correlation structure of the cell displacements
furthermore allowed us to perform a quantitative comparison of the cell trajectories from our model simulations with experimental data.
%
%We conclude that the displacement of amoeboid cell crawling and other types of motion like fan-shape cell motion follow basically the same mechanisms and it is only tuned by the values of certain parameters of the biochemical interactions and of the biophysical forces. 

Based on our simulations we furthermore conclude that a continuous transition between amoeboid and fan-shaped motions is a realistic scenario, as some of the predicted intermediate states observed in simulations have been confirmed in experiments with {\it D. discoideum} cells.
We speculate that the same model can be also employed to describe the motion of other cell types with different motion strategies such as keratocytes or fibroblasts.

%Finally using AR model we obtained the parameters that reproduces the tracks of the same type though a link between experimental and simulation results was not done.

\section*{Acknowledgment}
E.M. and C.B. acknowledge funding by the Deutsche Forschungsgemeinschaft in the framework of Sonderforschungsbereich 1294, project B02.
S.F. and C.B. gratefully acknowledge funding by the Deutsche Forschungsgemeinschaft in the framework of Sonderforschungsbereich 937, project A09. 
S.A. and E.M. thank support from MICINN (Spain), and FEDER (European Union), under project PGC2018-095456-B-I00. 
E.M. acknowledges also financial support from CONACYT. 
F.F. acknowledges financial support from the \it{Juan de la Cierva} programme (grant IJC2018-038463-I) from the Spanish MICINN, from the {\it Obra Social la Caixa} through the programme {\it Recerca en Matemàtica Col·laborativa} and the CERCA Programme of the {\it Generalitat de Catalunya}.
	
	\nocite{*} % to test all bib entrys
	%\bibliographystyle{unsrt} % <======================== not longer needed!
	%\bibliography{\jobname} % <========================== not longer needed!

\begin{thebibliography}{1} % <================================== mwe.bbl
		
	
		
	\bibitem{allard2013traveling}
	Jun Allard and Alex Mogilner.
	\newblock Traveling waves in actin dynamics and cell motility.
	\newblock {\em Current opinion in cell biology}, 25(1):107--115, 2013.
	
	\bibitem{Alonso18}
	Sergio Alonso, Maike Stange, and Carsten Beta.
	\newblock Modeling random crawling, membrane deformation and intracellular
	polarity of motile amoeboid cells.
	\newblock {\em PLOS ONE}, 13(8):1--22, 08 2018.
	
	\bibitem{altschuler2008spontaneous}
	Steven~J Altschuler, Sigurd~B Angenent, Yanqin Wang, and Lani~F Wu.
	\newblock On the spontaneous emergence of cell polarity.
	\newblock {\em Nature}, 454(7206):886, 2008.
	
	\bibitem{annesley_dictyostelium_2009}
	Sarah~J. Annesley and Paul~R. Fisher.
	\newblock Dictyostelium discoideum—a model for many reasons.
	\newblock {\em Molecular and Cellular Biochemistry}, 329(1-2):73--91, September
	2009.
	
	\bibitem{asano2004keratocyte}
	Yukako Asano, Takafumi Mizuno, Takahide Kon, Akira Nagasaki, Kazuo Sutoh, and
	Taro~QP Uyeda.
	\newblock Keratocyte-like locomotion in amib-null dictyostelium cells.
	\newblock {\em Cell motility and the cytoskeleton}, 59(1):17--27, 2004.
	
	\bibitem{Barnhart10}
	Erin~L. Barnhart, Greg~M. Allen, Frank J{\"u}licher, and Julie~A. Theriot.
	\newblock Bipedal locomotion in crawling cells.
	\newblock {\em Biophysical Journal}, 98(6):933 -- 942, 2010.
	
	\bibitem{beta2010bistability}
	Carsten Beta.
	\newblock Bistability in the actin cortex.
	\newblock {\em PMC biophysics}, 3(1):12, 2010.
	
	\bibitem{beta2008bistable}
	Carsten Beta, Gabriel Amselem, and Eberhard Bodenschatz.
	\newblock A bistable mechanism for directional sensing.
	\newblock {\em New Journal of Physics}, 10(8):083015, 2008.
	
	\bibitem{beta2017intracellular}
	Carsten Beta and Karsten Kruse.
	\newblock Intracellular oscillations and waves.
	\newblock {\em Annual Review of Condensed Matter Physics}, 8:239--264, 2017.
	
	\bibitem{bloomfield2015neurofibromin}
	Gareth Bloomfield, David Traynor, Sophia~P Sander, Douwe~M Veltman, Justin~A
	Pachebat, and Robert~R Kay.
	\newblock Neurofibromin controls macropinocytosis and phagocytosis in
	dictyostelium.
	\newblock {\em Elife}, 4:e04940, 2015.
	
	\bibitem{boettinger2002phase}
	William~J Boettinger, James~A Warren, Christoph Beckermann, and Alain Karma.
	\newblock Phase-field simulation of solidification.
	\newblock {\em Annual review of materials research}, 32(1):163--194, 2002.
	
	\bibitem{camley2013periodic}
	Brian~A Camley, Yanxiang Zhao, Bo~Li, Herbert Levine, and Wouter-Jan Rappel.
	\newblock Periodic migration in a physical model of cells on micropatterns.
	\newblock {\em Physical review letters}, 111(15):158102, 2013.
	
	\bibitem{Camley17}
	Brian~A. Camley, Yanxiang Zhao, Bo~Li, Herbert Levine, and Wouter-Jan Rappel.
	\newblock Crawling and turning in a minimal reaction-diffusion cell motility
	model: Coupling cell shape and biochemistry.
	\newblock {\em Phys. Rev. E}, 95:012401, Jan 2017.
	
	\bibitem{cao_minimal_2019}
	Yuansheng Cao, Elisabeth Ghabache, Yuchuan Miao, Cassandra Niman, Hiroyuki
	Hakozaki, Samara~L. Reck-Peterson, Peter~N. Devreotes, and Wouter-Jan Rappel.
	\newblock A minimal computational model for three-dimensional cell migration.
	\newblock {\em Journal of The Royal Society Interface}, 16(161):20190619,
	December 2019.
	
	\bibitem{cao2019plasticity}
	Yuansheng Cao, Elisabeth Ghabache, and Wouter-Jan Rappel.
	\newblock Plasticity of cell migration resulting from mechanochemical coupling.
	\newblock {\em Elife}, 8, 2019.
	
	\bibitem{cherstvy_non-gaussianity_2018}
	Andrey~G. Cherstvy, Oliver Nagel, Carsten Beta, and Ralf Metzler.
	\newblock Non-{Gaussianity}, population heterogeneity, and transient
	superdiffusion in the spreading dynamics of amoeboid cells.
	\newblock {\em Physical Chemistry Chemical Physics}, 20(35):23034--23054, 2018.
	
	\bibitem{collenburg2017activity}
	Lena Collenburg, Niklas Beyersdorf, Teresa Wiese, Christoph Arenz, Essa~M
	Saied, Katrin~Anne Becker-Flegler, Sibylle Schneider-Schaulies, and Elita
	Avota.
	\newblock The activity of the neutral sphingomyelinase is important in t cell
	recruitment and directional migration.
	\newblock {\em Frontiers in immunology}, 8:1007, 2017.
	
	\bibitem{edwards2018insight}
	Marc Edwards, Huaqing Cai, Bedri Abubaker-Sharif, Yu~Long, Thomas~J Lampert,
	and Peter~N Devreotes.
	\newblock Insight from the maximal activation of the signal transduction
	excitable network in dictyostelium discoideum.
	\newblock {\em Proceedings of the National Academy of Sciences},
	115(16):E3722--E3730, 2018.
	
	\bibitem{flemming2019}
	Sven Flemming, Francesc Font, Sergio Alonso, and Carsten Beta.
	\newblock How cortical waves drive fission of motile cells.
	\newblock {\em Proceedings of the National Academy of Sciences},
	117(12):6330--6338, 2020.
	
	\bibitem{folch1999phase}
	R~Folch, J~Casademunt, A~Hern{\'a}ndez-Machado, and L~Ramirez-Piscina.
	\newblock Phase-field model for hele-shaw flows with arbitrary viscosity
	contrast. i. theoretical approach.
	\newblock {\em Physical Review E}, 60(2):1724, 1999.
	
	\bibitem{frank2016frequent}
	Viktoria Frank, Stefan Kaufmann, Rebecca Wright, Patrick Horn, Hiroshi~Y
	Yoshikawa, Patrick Wuchter, Jeppe Madsen, Andrew~L Lewis, Steven~P Armes,
	Anthony~D Ho, et~al.
	\newblock Frequent mechanical stress suppresses proliferation of mesenchymal
	stem cells from human bone marrow without loss of multipotency.
	\newblock {\em Scientific reports}, 6(1):1--12, 2016.
	
	\bibitem{fukushima2019excitable}
	Seiya Fukushima, Satomi Matsuoka, and Masahiro Ueda.
	\newblock Excitable dynamics of ras triggers spontaneous symmetry breaking of
	pip3 signaling in motile cells.
	\newblock {\em J Cell Sci}, 132(5):jcs224121, 2019.
	
	\bibitem{Gerhardt4507}
	Matthias Gerhardt, Mary Ecke, Michael Walz, Andreas Stengl, Carsten Beta, and
	G{\"u}nther Gerisch.
	\newblock Actin and pip3 waves in giant cells reveal the inherent length scale
	of an excited state.
	\newblock {\em Journal of Cell Science}, 127(20):4507--4517, 2014.
	
	\bibitem{gerhardt_signaling_2014}
	Matthias Gerhardt, Michael Walz, and Carsten Beta.
	\newblock Signaling in chemotactic amoebae remains spatially confined to
	stimulated membrane regions.
	\newblock {\em J Cell Sci}, 127(23):5115--5125, December 2014.
	
	\bibitem{goehring2013cell}
	Nathan~W Goehring and Stephan~W Grill.
	\newblock Cell polarity: mechanochemical patterning.
	\newblock {\em Trends in cell biology}, 23(2):72--80, 2013.
	
	\bibitem{gorelik2014quantitative}
	Roman Gorelik and Alexis Gautreau.
	\newblock Quantitative and unbiased analysis of directional persistence in cell
	migration.
	\newblock {\em Nature protocols}, 9(8):1931, 2014.
	
	\bibitem{haastert_chemotaxis_2004}
	Peter J. M.~Van Haastert and Peter~N. Devreotes.
	\newblock Chemotaxis: signalling the way forward.
	\newblock {\em Nature Reviews Molecular Cell Biology}, 5(8):626--634, August
	2004.
	
	\bibitem{hiraiwa2014relevance}
	Tetsuya Hiraiwa, Akihiro Nagamatsu, Naohiro Akuzawa, Masatoshi Nishikawa, and
	Tatsuo Shibata.
	\newblock Relevance of intracellular polarity to accuracy of eukaryotic
	chemotaxis.
	\newblock {\em Physical biology}, 11(5):056002, 2014.
	
	\bibitem{iglesias2008navigating}
	Pablo~A Iglesias and Peter~N Devreotes.
	\newblock Navigating through models of chemotaxis.
	\newblock {\em Current opinion in cell biology}, 20(1):35--40, 2008.
	
	\bibitem{jilkine2011comparison}
	Alexandra Jilkine and Leah Edelstein-Keshet.
	\newblock A comparison of mathematical models for polarization of single
	eukaryotic cells in response to guided cues.
	\newblock {\em PLoS computational biology}, 7(4):e1001121, 2011.
	
	\bibitem{kockelkoren2003computational}
	Julien Kockelkoren, Herbert Levine, and Wouter-Jan Rappel.
	\newblock Computational approach for modeling intra-and extracellular dynamics.
	\newblock {\em Physical Review E}, 68(3):037702, 2003.
	
	\bibitem{kulawiak2016modeling}
	Dirk~Alexander Kulawiak, Brian~A Camley, and Wouter-Jan Rappel.
	\newblock Modeling contact inhibition of locomotion of colliding cells
	migrating on micropatterned substrates.
	\newblock {\em PLoS computational biology}, 12(12):e1005239, 2016.
	
	\bibitem{lee2014automated}
	Chen-Yu Lee, Sukryool Kang, Andrew~D Chisholm, and Pamela~C Cosman.
	\newblock Automated cell junction tracking with modified active contours guided
	by sift flow.
	\newblock In {\em 2014 IEEE 11th International Symposium on Biomedical Imaging
		(ISBI)}, pages 290--293. IEEE, 2014.
	
	\bibitem{lober2014modeling}
	Jakob L{\"o}ber, Falko Ziebert, and Igor~S Aranson.
	\newblock Modeling crawling cell movement on soft engineered substrates.
	\newblock {\em Soft matter}, 10(9):1365--1373, 2014.
	
	\bibitem{lober_collisions_2015}
	Jakob L{\"o}ber, Falko Ziebert, and Igor~S. Aranson.
	\newblock Collisions of deformable cells lead to collective migration.
	\newblock {\em Scientific Reports}, 5:9172, March 2015.
	
	\bibitem{lustig2019noninvasive}
	Maayan Lustig, Qingling Feng, Yohan Payan, Amit Gefen, and Dafna Benayahu.
	\newblock Noninvasive continuous monitoring of adipocyte differentiation: From
	macro to micro scales.
	\newblock {\em Microscopy and Microanalysis}, 25(1):119--128, 2019.
	
	\bibitem{makarava_quantifying_2014}
	Natallia Makarava, Stephan Menz, Matthias Theves, Wilhelm Huisinga, Carsten
	Beta, and Matthias Holschneider.
	\newblock Quantifying the degree of persistence in random amoeboid motion based
	on the {Hurst} exponent of fractional {Brownian} motion.
	\newblock {\em Physical Review E}, 90(4):042703, October 2014.
	
	\bibitem{maree2006polarization}
	Athanasius~FM Mar{\'e}e, Alexandra Jilkine, Adriana Dawes, Ver{\^o}nica~A
	Grieneisen, and Leah Edelstein-Keshet.
	\newblock Polarization and movement of keratocytes: a multiscale modelling
	approach.
	\newblock {\em Bulletin of mathematical biology}, 68(5):1169--1211, 2006.
	
	\bibitem{matsuoka2018mutual}
	Satomi Matsuoka and Masahiro Ueda.
	\newblock Mutual inhibition between pten and pip3 generates bistability for
	polarity in motile cells.
	\newblock {\em Nature communications}, 9(1):4481, 2018.
	
	\bibitem{miao2017altering}
	Yuchuan Miao, Sayak Bhattacharya, Marc Edwards, Huaqing Cai, Takanari Inoue,
	Pablo~A Iglesias, and Peter~N Devreotes.
	\newblock Altering the threshold of an excitable signal transduction network
	changes cell migratory modes.
	\newblock {\em Nature cell biology}, 19(4):329, 2017.
	
	\bibitem{mogilner2012cell}
	Alex Mogilner, Jun Allard, and Roy Wollman.
	\newblock Cell polarity: quantitative modeling as a tool in cell biology.
	\newblock {\em Science}, 336(6078):175--179, 2012.
	
	\bibitem{mogilner_experiment_2019}
	Alex Mogilner, Erin~L. Barnhart, and Kinneret Keren.
	\newblock Experiment, theory, and the keratocyte: {An} ode to a simple model
	for cell motility.
	\newblock {\em Seminars in Cell \& Developmental Biology}, page
	S1084952119300369, November 2019.
	
	\bibitem{mori2008wave}
	Yoichiro Mori, Alexandra Jilkine, and Leah Edelstein-Keshet.
	\newblock Wave-pinning and cell polarity from a bistable reaction-diffusion
	system.
	\newblock {\em Biophysical journal}, 94(9):3684--3697, 2008.
	
	\bibitem{moure2016computational}
	Adrian Moure and Hector Gomez.
	\newblock Computational model for amoeboid motion: Coupling membrane and
	cytosol dynamics.
	\newblock {\em Physical Review E}, 94(4):042423, 2016.
	
	\bibitem{moure2017phase}
	Adrian Moure and Hector Gomez.
	\newblock Phase-field model of cellular migration: Three-dimensional
	simulations in fibrous networks.
	\newblock {\em Computer Methods in Applied Mechanics and Engineering},
	320:162--197, 2017.
	
	\bibitem{nagel_geometry-driven_2014}
	Oliver Nagel, Can Guven, Matthias Theves, Meghan Driscoll, Wolfgang Losert, and
	Carsten Beta.
	\newblock Geometry-{Driven} {Polarity} in {Motile} {Amoeboid} {Cells}.
	\newblock {\em PLoS ONE}, 9(12):e113382, December 2014.
	
	\bibitem{najem2013phase}
	Sara Najem and Martin Grant.
	\newblock Phase-field approach to chemotactic driving of neutrophil
	morphodynamics.
	\newblock {\em Physical Review E}, 88(3):034702, 2013.
	
	\bibitem{Nickaeen17}
	Masoud Nickaeen, Igor~L. Novak, Stephanie Pulford, Aaron Rumack, Jamie Brandon,
	Boris~M. Slepchenko, and Alex Mogilner.
	\newblock A free-boundary model of a motile cell explains turning behavior.
	\newblock {\em PLOS Computational Biology}, 13(11):1--22, 11 2017.
	
	\bibitem{otsuji2007mass}
	Mikiya Otsuji, Shuji Ishihara, Kozo Kaibuchi, Atsushi Mochizuki, Shinya Kuroda,
	et~al.
	\newblock A mass conserved reaction--diffusion system captures properties of
	cell polarity.
	\newblock {\em PLoS computational biology}, 3(6):e108, 2007.
	
	\bibitem{paschke2018rapid}
	Peggy Paschke, David~A Knecht, Augustinas Silale, David Traynor, Thomas~D
	Williams, Peter~A Thomason, Robert~H Insall, Jonathan~R Chubb, Robert~R Kay,
	and Douwe~M Veltman.
	\newblock Rapid and efficient genetic engineering of both wild type and axenic
	strains of dictyostelium discoideum.
	\newblock {\em PLoS One}, 13(5):e0196809, 2018.
	
	\bibitem{pons2010helical}
	Antonio~J Pons and Alain Karma.
	\newblock Helical crack-front instability in mixed-mode fracture.
	\newblock {\em Nature}, 464(7285):85--89, 2010.
	
	\bibitem{rappel2017mechanisms}
	Wouter-Jan Rappel and Leah Edelstein-Keshet.
	\newblock Mechanisms of cell polarization.
	\newblock {\em Current opinion in systems biology}, 3:43--53, 2017.
	
	\bibitem{schroth-diez_propagating_2009}
	Britta Schroth-Diez, Silke Gerwig, Mary Ecke, Reiner Hegerl, Stefan Diez, and
	Günther Gerisch.
	\newblock Propagating waves separate two states of actin organization in living
	cells.
	\newblock {\em HFSP Journal}, 3(6):412--427, December 2009.
	
	\bibitem{shao2012coupling}
	Danying Shao, Herbert Levine, and Wouter-Jan Rappel.
	\newblock Coupling actin flow, adhesion, and morphology in a computational cell
	motility model.
	\newblock {\em Proceedings of the National Academy of Sciences},
	109(18):6851--6856, 2012.
	
	\bibitem{shao2010computational}
	Danying Shao, Wouter-Jan Rappel, and Herbert Levine.
	\newblock Computational model for cell morphodynamics.
	\newblock {\em Physical review letters}, 105(10):108104, 2010.
	
	\bibitem{swaney2010eukaryotic}
	Kristen~F Swaney, Chuan-Hsiang Huang, and Peter~N Devreotes.
	\newblock Eukaryotic chemotaxis: a network of signaling pathways controls
	motility, directional sensing, and polarity.
	\newblock {\em Annual review of biophysics}, 39:265--289, 2010.
	
	\bibitem{takagi2008functional}
	Hiroaki Takagi, Masayuki~J Sato, Toshio Yanagida, and Masahiro Ueda.
	\newblock Functional analysis of spontaneous cell movement under different
	physiological conditions.
	\newblock {\em PloS one}, 3(7), 2008.
	
	\bibitem{taniguchi2013phase}
	Daisuke Taniguchi, Shuji Ishihara, Takehiko Oonuki, Mai Honda-Kitahara,
	Kunihiko Kaneko, and Satoshi Sawai.
	\newblock Phase geometries of two-dimensional excitable waves govern
	self-organized morphodynamics of amoeboid cells.
	\newblock {\em Proceedings of the National Academy of Sciences},
	110(13):5016--5021, 2013.
	
	\bibitem{van2017coupled}
	Peter~JM van Haastert, Ineke Keizer-Gunnink, and Arjan Kortholt.
	\newblock Coupled excitable ras and f-actin activation mediates spontaneous
	pseudopod formation and directed cell movement.
	\newblock {\em Molecular biology of the cell}, 28(7):922--934, 2017.
	
	\bibitem{winkler_confinement_2019}
	Benjamin Winkler, Igor~S. Aranson, and Falko Ziebert.
	\newblock Confinement and substrate topography control cell migration in a {3D}
	computational model.
	\newblock {\em Communications Physics}, 2(1):1--11, July 2019.
	
	\bibitem{ziebert2011model}
	Falko Ziebert, Sumanth Swaminathan, and Igor~S Aranson.
	\newblock Model for self-polarization and motility of keratocyte fragments.
	\newblock {\em Journal of The Royal Society Interface}, 9(70):1084--1092, 2011.

		
	\end{thebibliography}
	 % <======================================= mwe.bbl
	
\end{document}